\documentclass[aps,pre,showpacs,twocolumn]{revtex4}
\usepackage{amsmath}
\usepackage{amsfonts}
\usepackage{graphicx}
\usepackage{overpic}
\usepackage{psfrag}
\usepackage{color}

\begin{document}

\title{Free energy of colloidal particles at the surface of sessile drops}
\author{J.~Guzowski}
\author{M.~Tasinkevych}
\author{S.~Dietrich}
\affiliation{Max--Planck--Institut f\"ur Metallforschung,
Heisenbergstr.\ 3, 70569 Stuttgart, Germany\\
and\\
Institut f\"ur Theoretische und Angewandte Physik, Universit\"at Stuttgart,
 Pfaffenwaldring 57, 70569 Stuttgart, Germany}
\date{\today}

\begin{abstract}
The influence of finite system size on the free energy of a spherical particle floating at the surface of a sessile droplet is studied both analytically and numerically. In the special case that the contact angle at the substrate equals $\pi/2$ a capillary analogue of the method of images is applied in order to calculate small deformations of the droplet shape if an external force is applied to the particle. The type of boundary conditions for the droplet shape at the substrate determines the sign of the capillary monopole associated with the image particle. Therefore, the free energy of the particle, which is proportional to the interaction energy of the original particle with its image, can be of either sign, too. The analytic solutions, given by the Green's function of the capillary equation, are constructed such that the condition of the forces acting on the droplet being balanced and of the volume constraint are fulfilled. Besides the known phenomena of attraction of a particle to a free contact line and repulsion from a pinned one, we observe a local free energy minimum for the particle being located at the drop apex or at an intermediate angle, respectively. This peculiarity can be traced back to a non-monotonic behavior of the Green's function, which reflects the interplay between the deformations of the droplet shape and the volume constraint.
\end{abstract}


\maketitle

\section{Introduction.}

Two-dimensional (2D) structures formed by micron-sized colloidal particles trapped at fluid interfaces have received a significant attention since an interfacial colloidal crystal has been first observed by Pieranski in 1980~\cite{Pieranski1980}. In the last two decades 2D-colloidal structures have proven to be an excellent model system for studying the influences of dimensionality and confinement on  phase transitions. For example, a general scenario for 2D-melting, predicted by Kosterlitz and Thouless~\cite{Kosterlitz1973} in 1973, has been directly verified in experiments using quasi-2D colloidal structures~\cite{Zahn2000}. It has also been found that strong lateral confinement can lead to peculiarities like reentrant freezing~\cite{Bubeck1999}. From the theoretical point of view the capillary interactions between colloidal particles mediated by a fluid interface resemble 2D-electrostatics, which is due to the fact that the small deformations of the interface are governed by the Poisson equation with the pressure field playing the role of a two-dimensional charge density.
%This suggests the possibility of using colloids in studying two-dimensional gravitating systems. 
In terms of application perspectives, controlling self-assembly~\cite{Bowden1997} and structure formation~\cite{Helseth2005,Aizenberg2000,Loudet2009} at fluid interfaces provides tools for the engineering of advanced materials based on transferring such microstructures onto solid substrates~\cite{Aizenberg2000}. Colloidal particles are also known to stabilize emulsions by self-assembling at the droplet surfaces~\cite{Pickering1907}, creating capsules which might have potential use in medicine~\cite{Dinsmore2002}. The particles at droplet surfaces can also be used to study correlations~\cite{Chavez-Paez2003,Viveros-Mendez2008}, topological defects, and the generalized Thomson problem for optimal packing on a sphere~\cite{Bausch2003}. 

The presence of an interface strongly modifies the effective interactions between colloidal particles as compared with their counterparts in the bulk. It has been already noted by Pieranski~\cite{Pieranski1980}, that in the generic case of different dielectric constants of the adjacent fluids, the interface breaks the symmetry in the ion distribution around each particle, leading to an effective dipolar repulsion. Therefore, in the presence of lateral confinement a 2D-crystal can form. However, in the last decade many authors reported on the spontaneous formation of mesostructures like clusters, stripes~\cite{Ruiz-Garcia1998,Ghezzi1997,Ghezzi2001,Sear1999}, and freely floating crystallites~\cite{Nikolaides2002} without confinement. These observations suggest more complicated forms of the effective interaction potentials, in particular the presence of an attraction, the origin of which has recently been intensively discussed in the literature. The interface mediated capillary interactions are considered to be promising candidates for an explanation, but their theoretical description is still in progress. The main idea, formulated already in 1949 by Nicolson~\cite{Nicolson1949}, is that the surface energy of a deformed interface depends on the separation of the particles generating the deformation~\cite{Kralchevsky1992}. Originally, Nicolson studied the behavior of millimeter-sized bubbles deforming the interface due to their buoyancy. In the case of micrometer-sized or smaller particles gravity can be usually neglected, but even then capillary interactions can dominate thermal energies by orders of magnitude~\cite{Stamou2000,Loudet2005}. There seem to be at least two reasons for the origin of the interface deformation around the particles. First, in the case of charged colloidal particles pinned at the interface, an electrostatic pressure emerges due to the inhomogeneous ion distribution close to the interface~\cite{Oettel2005,Dominguez2007a}. Second, the interface must deform in order to meet the particle surface at a given contact angle or at a prescribed (pinned) contact line, which becomes important in the cases of smooth non-spherical particles~\cite{Loudet2005,Lehle2008} or rough particles~\cite{Stamou2000}, respectively. 

The case of charged particles has been strongly debated in the last few years and different theoretical approaches have lead to contradictory results for the form of the capillary interaction potential. In 2002 Nikolaides \textit{et\ al.}~\cite{Nikolaides2002} observed a freely floating hexagonal structure formed by charged particles at the surface of a water droplet in oil. The authors proposed a simple explanation according to which the particles are pulled towards the water phase by the electrostatic forces due to the different dielectric constants of water and oil. They concluded that the presence of such a pulling force leads to a logarithmically varying  interface profile around a single particle and consequently to a logarithmically varying effective interaction potential between two particles. However, it was noted by Megens \textit{et\ al.}~\cite{Megens2003} that this long-ranged interface deformation yields an unbalanced force acting on the rim of the vessel containing the liquid, which is inconsistent with Newton's third law. Instead, in the presence of the electrostatic pressure, decaying as $r^{-6}$ with lateral distance $r$ from the particle, the latter authors concluded that the interaction potential should be attractive and should decay also like $r^{-6}$. A similar reasoning was presented by Foret and W\"urger~\cite{Foret2004}, who obtained the same power law, but with opposite sign. Their calculation was based on a regularization of the interface deformation field at $r=0$ and a superposition approximation for the electrostatic pressure in the presence of two particles. However, Dom\'{\i}nguez \textit{et al.}\ noted~\cite{Dominguez2007a} that the latter assumption is not correct because the electrostatic pressure, proportional to the electrostatic field squared, is not additive. Instead, one should use the additivity of the electrostatic potential, which finally gives an attractive $r^{-3}$ behavior. This leads to the conclusion, that the total interparticle potential including the direct dipolar electrostatic repulsion, governed also by the power law $r^{-3}$, cannot exhibit a minimum, unless the screening length in water is comparable with the size of the particles. In such a case a shallow minimum with a depth of several $k_B T$ could be expected~\cite{Dominguez2007a}. However, its occurrence would depend sensitively on the precise values of several system parameters. 

In yet another publication~\cite{Dominguez2005}, Dom\'{\i}nguez \textit{et al.}\ suggested that the finite size of the system might be important for generating the long-ranged interactions. In fact, in the experiment reported by Nikolaides \textit{et al.}, the particles were trapped at the surface of a droplet pinned to a solid plate, but curvature and volume constraint were neglected in the theoretical description in Ref.~\cite{Nikolaides2002}. For recent assessments of these capillary forces see Refs.~\cite{Danov2010}, ~\cite{Oettel2008}, and \cite{Dominguez2008a}. In the present work we show that finite size effects indeed lead to effects not present in the case of a flat unbounded interface. Quite generally, the presence of boundaries in a colloidal system can change the nature of the effective interactions. For example, the hydrodynamic interactions in the bulk decay like the inverse of the particle separation, but in the presence of one or two walls or of surfaces the decay exponent can change or there might be even a switch to a logarithmic law~\cite{Diamant2009}. On the other hand, in one-dimensional channels the hydrodynamic interactions are screened and characterized by an exponential decay~\cite{Cui2002}. Our ultimate objective is to study how curvature and confinement of the fluid interface affect the capillary interactions between the floating particles. The effects of curvature have been already studied by W\"urger~\cite{Wurger2006,Wurger2006a}, but this description cannot be regarded as complete as long as the consequences of force balance on the droplet have not been correctly resolved~\cite{Dominguez2007}. In Ref.~\cite{Kralchevsky1995} Kralchevsky \textit{et al.}\ have provided numerical results for the free energy of particles protruding from a spherical liquid film and in some cases they observed a non-monotonic dependence of the ensuing effective interaction on the spatial distance between them, but the physical mechanism responsible for this effect has not been explained. Oettel \textit{et al.}~\cite{Oettel2005} derived the equation for the axisymmetric shape of a slightly deformed spherical droplet, taking into account the volume constraint and a certain type of boundary condition at the substrate. However, the influence of the type of boundary condition on the free energy and the occurrence of a possible confining capillary potential due to the presence of the substrate have not yet been discussed. In the following we aim at calculating such an effective potential and at providing some general instructions of how the finite size effects should be taken into account when discussing capillary interactions.

As a first step in understanding the interactions, we have performed a detailed analysis of the one-body problem of a spherical particle at the surface of a sessile droplet (Sec.\ II). In Sects. III and IV we propose a perturbation theory for a slightly deformed droplet and solve the boundary value problem at the substrate by the method of images known from electrostatics. In Sec.\ V we present the results of the full numerical minimization of the free energy and compare them with the analytical results. Together with our conclusions we summarize our findings in Sec.\ VI. Various important analytic calculations are presented in Appendices A, B, and C.

%%%%%%%%%%%%%%%%%%%%%%%%%%%%%%%%%%%%%%%%%%%%%%%%%%%%%%%%%%%%%%%%%%%%%%%%%%%%%%%%%%%%%
%%%%%%%%%%%%%%%%%%%%%%%%%%%%%%%%%%%%%%%%%%%%%%%%%%%%%%%%%%%%%%%%%%%%%%%%%%%%%%%%%%%%%

\section{Model} 
\begin{figure}[ht]
	\centering
	\psfragscanon
	\psfrag{t0}[c][c][1]{$\theta_0$}
	\psfrag{r0}[c][c][1]{$R_0\;$}
	\psfrag{a}[c][c][1]{$\alpha$}
	\psfrag{2a}[c][c][1]{$2a$}
	\psfrag{x}[c][c][1]{$\,x$}
	\psfrag{y}[c][c][1]{$\,y$}
	\psfrag{z}[c][c][1]{$z\,$}
	\psfrag{tp}[c][c][1]{$\theta_p$}
	\psfrag{r1}[c][c][1]{$\!\boldsymbol{r}_{1,ref}$}
	\begin{overpic}[width=0.5\textwidth]{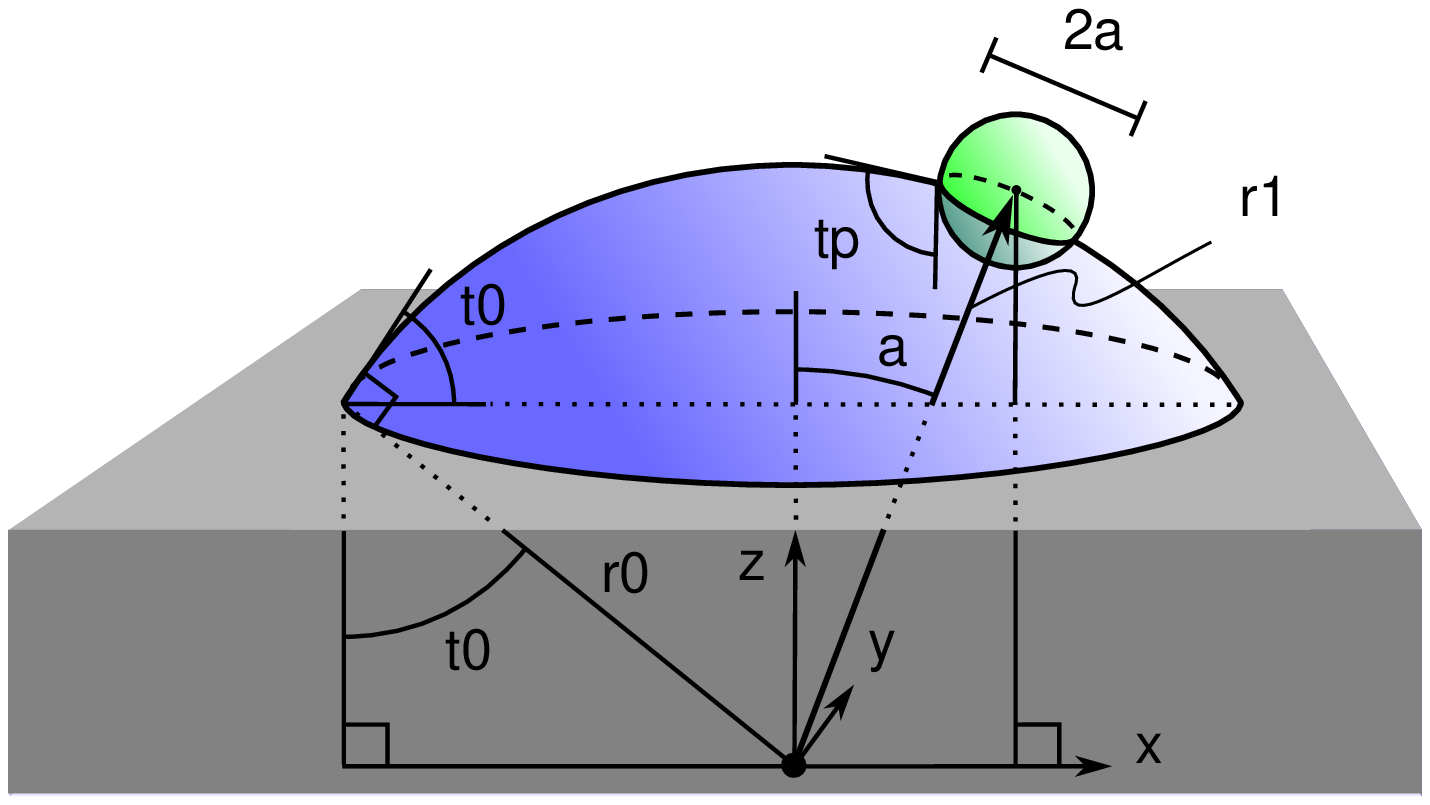}
	\put(5,40){\begin{large}$(a)$\end{large}}
	\end{overpic}\\
	\psfragscanoff
	\vspace*{0.5cm}
	\psfragscanon
	\psfrag{t0}[c][c][1]{$\theta_0$}
	\psfrag{tc}[c][c][1]{$\theta_c(\phi)$}
	\psfrag{r0}[c][c][1]{$R_0$}
	\psfrag{a}[c][c][1]{$\alpha$}
	\psfrag{x}[c][c][1]{$x$}
	\psfrag{y}[c][c][1]{$y$}
	\psfrag{z}[c][c][1]{$z$}
	\psfrag{tp}[c][c][1]{$\theta_p$}
	\psfrag{ph}[c][c][1]{$\phi$}
	\psfrag{f}[c][c][1]{$\boldsymbol{f}$}
	\psfrag{rh}[c][c][1]{$\quad\boldsymbol{r}_{1,ref}+h\boldsymbol{e}_r$}
	\begin{overpic}[width=0.5\textwidth]{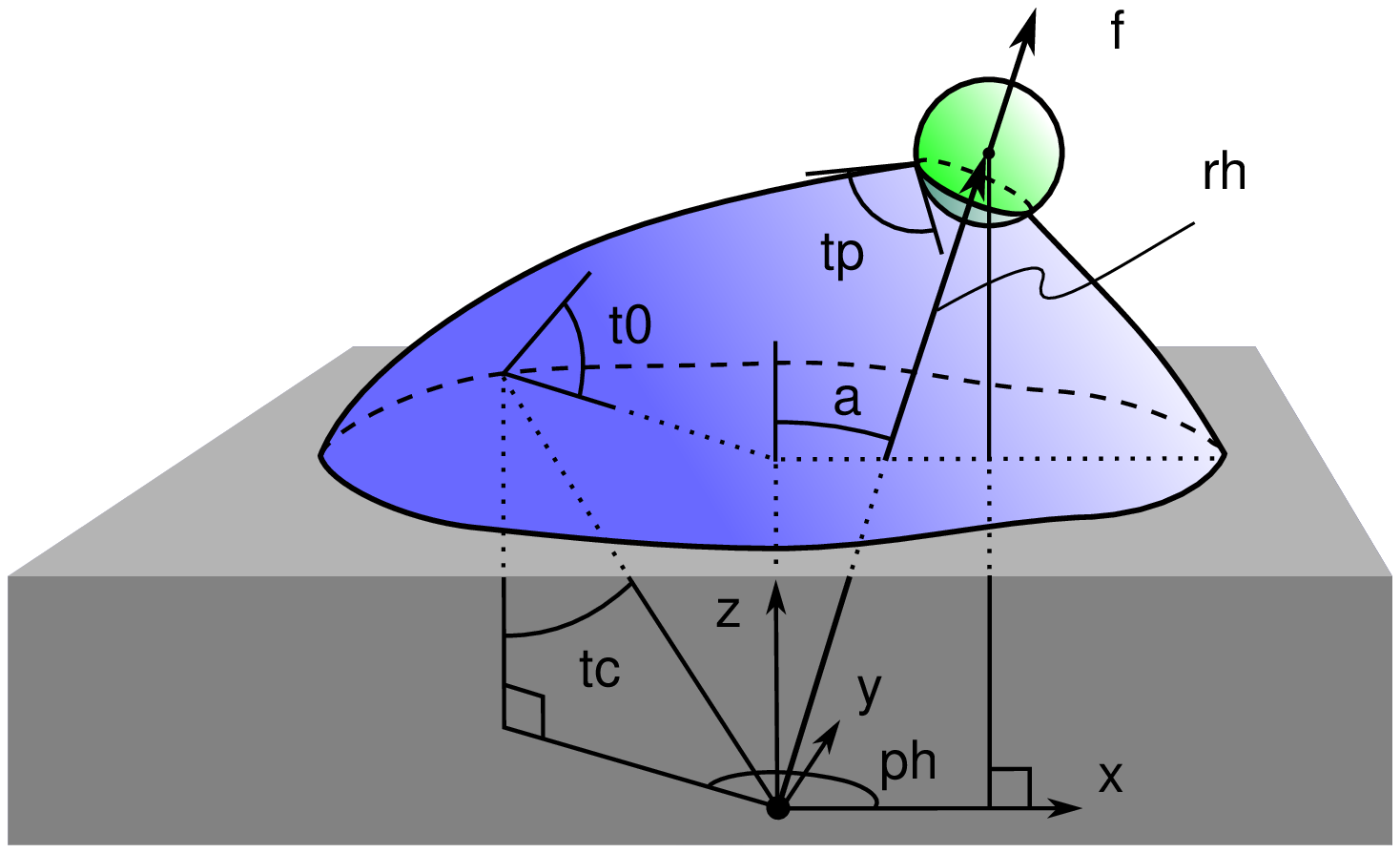}
	\put(5,40){\begin{large}$(b)$\end{large}}
	\end{overpic}
	\psfragscanoff
	\caption{Sketch of the system (see main text); $(a)$ reference configuration; $(b)$ deformation due to an external force $\boldsymbol{f}=f\boldsymbol{e}_r$ for a free contact line at the substrate, which is not necessarily a circle but it is characterized by the angle $\theta_c(\phi)$ as a function of the azimuthal angle $\phi$ in the $xy$ plane. Neglecting the line tension, in this case the contact angle at the substrate equals $\theta_0$ along the contact line. The direction of the $x$-axis is chosen such that the center of the particle at $\boldsymbol{r}_{1,ref}+h\boldsymbol{e}_r$ (with $h=0$ in $(a)$) lies in the $xz$ plane. Accordingly, the external force acts in the direction $\Omega_1=(\theta=\alpha,\phi=0)$. In the case of a pinned contact line at the substrate (not shown), for $f\neq0$ and $\alpha\neq0$ the contact angle at the substrate varies along the contact line (i.e., it is a function of $\phi$) even if the contact line forms a circle. The contact line at the surface of the colloidal particle is taken to be free such that the corresponding contact angle $\theta_p$ does not vary along that contact line.}
	\label{sketch}
\end{figure}

We consider a smooth, solid spherical particle of radius $a$ trapped at the surface of a liquid droplet of volume $V_l$ and surface tension $\gamma$ residing on a planar substrate. For convenience we call the surrounding medium "gas", but we assume a non-volatile liquid. The equilibrium contact angles $\theta_0$ at the substrate and $\theta_p$ at the particle are given by Young's law:
\begin{align}
\cos\theta_0 &= \frac{\gamma_{0g}-\gamma_{0l}}{\gamma}
\label{young_0},\\
\cos\theta_p &= \frac{\gamma_{pg}-\gamma_{pl}}{\gamma}
\label{young_p},
\end{align}

\noindent where $\gamma_{ab}$ are surface tensions with the indices $a,b=0,p,l,g$ standing for substrate, particle, liquid, and gas, respectively, and $\gamma_{lg}\equiv\gamma$.
In polar coordinates, the position of the particle can be described by the radial displacement $h$ and the polar angle $\alpha$ (see Fig.~\ref{sketch}).

It can be shown~\cite{Kralchevsky_book} that in the absence of external forces the particle adjusts its immersion, independently of its lateral position at the droplet, according to Young's law (Eq.~(\ref{young_p})) such that the droplet remains an undeformed spherical cap. This configuration with the particle at the unperturbed droplet will be called the reference configuration, for which we set $h=0$. As a consequence of the spherical shape of the droplet the corresponding free energy does not depend on the polar angle $\alpha$ (see Fig.~\ref{sketch}). The only exceptions are configurations with a particle close to the contact line, where the movement of the particle is constrained by the substrate. This situation is common in experiments with evaporating droplets due to the flux of liquid dragging the particles towards the contact line~\cite{Sangani2009}. When the height of the interface in the liquid wedge reduces below the particle diameter the interface around the particle must deform. This particular case could be discussed within the framework presented here, too, but it is beyond the scope of the present study.

\subsection{Free energy functional}
If, however, an external force $f$ acts on the particle in radial direction, the interface deforms such that the capillary force counterbalances the external force. The corresponding equilibrium shape of the droplet minimizes the following free energy functional: 
 \begin{multline}
%\begin{equation}
	\mathcal{F}[\{\boldsymbol{r}(\Omega)\},h,\alpha;f,\{\gamma_{ab}\},a,V_l,\lambda] =\\
	= \gamma (S_{lg}-S_{lg,ref}) +(\gamma_{0l}-\gamma_{0g})(S_{0l}-S_{0l,ref}) \\
	+ (\gamma_{pl}-\gamma_{pg})(S_{pl}-S_{pl,ref})
	 - fh - \lambda(V-V_l),
	\label{functional}
%\end{equation}
\end{multline}

\noindent where $\{\boldsymbol{r}(\Omega)\}$ denotes the configuration of the interface in terms of spherical coordinates $\Omega=(\theta,\phi)$ on the unit sphere, and $S_{lg},S_{0l},S_{pl}$ are the liquid-gas, substrate-liquid, and particle-liquid surface areas, respectively. In Eq.\ (\ref{functional}) $\mathcal F$ is the free energy functional relative to the reference configuration of the unperturbed droplet referred to by the index $ref$. The last term with the Lagrange multiplier $\lambda$ ensures that the liquid volume is constant. The sign of this term has been chosen such that the value of $\lambda$ minimizing the functional in Eq.\ (\ref{functional}) can be interpreted as the internal pressure. The sizes of the droplet ($R_0$) and of the particle ($a$) are considered to be sufficiently large so that in Eq.\ (\ref{functional}) we can neglect line tension contributions and Eqs.\ (\ref{young_0}) and (\ref{young_p}) do indeed hold~\cite{Schimmele2007}.

The condition of mechanical equilibrium of the droplet in the presence of an external force requires a mechanism fixing the overall lateral position of the droplet on the substrate. The substrate provides a counterbalancing force in the vertical direction, but not in the lateral direction, and therefore the droplet would start to move under the action of the external force. There is motionless equilibrium only if either a body-force fixes the center of mass of the droplet or, alternatively, the contact line at the substrate is pinned. Both situations can occur in experiments. For example, in the presence of gravity a small tilt of the substrate could provide a lateral body-force counterbalancing the lateral component of $f\boldsymbol{e}_r$. (However, in this case also the weight of the fluid must be taken into account.) On the other hand, pinning of the contact line can occur due to heterogeneities of the substrate. To some extent such a pinning is always present in actual systems because the substrate is never perfectly smooth. Therefore, in practical terms, we expect this latter case to be the more relevant one. These two different conditions for mechanical equilibrium impose distinct additional constraints onto the equilibrium shape of the droplet and thus onto the minimization of the free energy functional in Eq.\ (\ref{functional}). We shall use an index $\sigma$ in order to distinguish between the cases of a free $(\sigma=A)$ or a pinned $(\sigma=B)$ contact line, respectively. In the limit of small deformations the forthcoming detailed analysis will reveal that those two cases correspond to Robin and Dirichlet boundary conditions, respectively.

A fixed lateral position $x_{CM}$ of the center of mass $(CM)$ can be achieved by adding a term $-f_{CM}(x_{CM}-x_{CM,ref})$ to the free energy functional $\mathcal{F}$, where $x_{CM,ref}$ is determined by the reference configuration and $-f_{CM}$ is a Lagrange multiplier. Choosing the minus sign in the prefactor $-f_{CM}$ will enable us to identify $f_{CM}$ with the force acting at the center of mass and fixing it. We introduce the free energy $F_A^{\star}$ as a constrained minimum of $\mathcal{F}$, with $\alpha$ kept constant and with a fixed center of mass:
%\begin{equation}
\begin{multline}
	F_A^{\star}(\alpha;f,\theta_0,\theta_p,a,R_0,\lambda,f_{CM}) =\\
	= \min_{\{\boldsymbol{r}(\Omega)\},h}\, \big[\mathcal{F}-f_{CM}(x_{CM}-x_{CM,ref})\big],
	\label{minimizeN}
\end{multline}
%\end{equation}

\noindent where the contact angles $\theta_0$ and $\theta_p$ are given by Eqs.\ (\ref{young_0}) and (\ref{young_p}), respectively, and $R_0=R_0(\theta_0,\theta_p,a,V_l)$ is the radius of the droplet in the reference configuration. In the following, if not indicated otherwise, we shall suppress the explicit dependence on the set $\{f,\theta_0,\theta_p,a,R_0\}$ of independent system parameters. The Lagrange multipliers $\lambda$ and $f_{CM}$ can be determined as functions of these independent system parameters from the conditions
\begin{align}
	V(\alpha;\lambda,f_{CM})&=V_l, \label{vol_constr}\\
	x_{CM}(\alpha;\lambda,f_{CM})&=x_{CM,ref}\label{xcm_constr}.
\end{align}

\noindent This renders $\lambda=\lambda(\alpha;f,\theta_0,\theta_p,a,R_0)$ and $f_{CM}(\alpha;f,\theta_0,\theta_p,a,R_0)$ which upon insertion into $F_A^{\star}(\alpha;f,\theta_0,\theta_p,a,R_0,\lambda,f_{CM})$ yields $F_A(\alpha;f,\theta_0,\theta_p,a,R_0)$.

On the other hand, the condition of a pinned contact line imposes only a geometric constraint onto the minimization of the free energy, which we symbolically indicate as
\begin{equation}
	F_B^{\star}(\alpha) = {\min_{\{\boldsymbol{r}(\Omega)\},h}}\!\!\!' \;\; \mathcal{F},
	\label{minimizeD}
\end{equation}

\noindent where $\min'$ indicates minimizing only over those configurations for which the contact line at the substrate lies at a circle corresponding to the one of the reference configuration (Fig.\ \ref{sketch}$(a)$). Inserting $\lambda$ and $f_{CM}$ from Eqs.\ (\ref{vol_constr}) and (\ref{xcm_constr}) into $F_B^{\star}$ yields $F_B(\alpha;f,\theta_0,\theta_p,a,R_0)$.

In both cases one can split the free energy into two parts:
\begin{equation}
	F_{\sigma}(\alpha) = F_{\sigma0} + \Delta F_{\sigma}(\alpha),
	\label{excess_free_en}
\end{equation}

\noindent where $\sigma=A,B$. $F_{\sigma0}:=F_{\sigma}(\alpha=0)$ is the free energy of the droplet with the adsorbed particle in the axially symmetric position $\alpha=0$ and $\Delta F_{\sigma} := F_{\sigma} - F_{\sigma0}$ is the excess free energy depending on the angular deviation of the particle position from the symmetry axis. The latter quantity will be the main focus of the following analysis.

%%%%%%%%%%%%%%%%%%%%%%%%%%%%%%%%%%%%%%%%%%%%%%%%%%%%%%%%%%%%%%%%%%%%%%%%%%%%%%%%%%%%%
%%%%%%%%%%%%%%%%%%%%%%%%%%%%%%%%%%%%%%%%%%%%%%%%%%%%%%%%%%%%%%%%%%%%%%%%%%%%%%%%%%%%%

\subsection{Parameterization in terms of spherical coordinates and effective description of the particle}
If without a particle the original fluid interface is flat, the capillary deformation of the interface around an added particle can be described in terms of capillary multipoles, uniquely determined by the deformation of the fluid interface outside the particle. Effectively, the interface with a particle can be replaced by the interface without the particle, but with an infinite set of point-multipoles, all centered at a single point inside the region originally occupied by the particle~\cite{Dominguez_book}. In the following analysis we apply a similar approach for the case of a spherical interface. The subsequent comparison with full numerical results will show that in the model studied here the effect of the particle onto the interface can be very well approximated by a capillary monopole. Nevertheless, for the present curved interface, we do not claim that there exists an equivalent set of point-multipoles positioned at a single point, analogous to the case of a flat interface, which is determined uniquely by the deformation of the interface outside the particle. Actually, in general, owing to the volume constraint, this might not be the case~\cite{Dominguez_private}. Instead, we assume that the particle can be replaced by a set of capillary charges distributed over the additional virtual piece of liquid-gas interface inside the particle. We introduce a pressure field $\Pi(\Omega)$ defined at the virtual piece of the interface described by the solid angle $\Delta\Omega$. (In the reference configuration, this piece of interface, defined at $\Delta\Omega_{ref}$, together with the remaining part of the interface, forms a perfect cap of a sphere.) Within this approach the contact line at the particle surface is virtual and enters only through $\Pi(\Omega)$; therefore, in the following, we shall use the notion ''contact line'' exclusively for the contact line at the substrate. In polar coordinates, one can express the free energy functional in Eq.\ (\ref{functional}) as a functional of the radial deformation $u(\Omega)=r(\Omega)-R_0$:
\begin{multline}
 	\mathcal{F}[\{u(\Omega)\}] = \gamma\int_{\Omega_c}\! d\Omega\, \big[s(u,\nabla_a u)-R_0^2\big]\\
 	+\gamma R_0^2 \left(\int_{\Omega_c\setminus\Omega_0}\! d\Omega\, -\int_{\Omega_0\setminus\Omega_c}\! d\Omega\right)\\
	-\dfrac{\gamma\cos\theta_0}{2}\int_0^{2\pi}\!d\phi\, \big[(R_0+u_c(\phi))^2\sin^2\theta_c(\phi)-R_0^2\sin^2\theta_0\big]\\	
	-\dfrac{1}{3}\int_{\Delta\Omega}\! d\Omega\,\Pi(\Omega)\big[(R_0+u)^3-R_0^3\big]\\
	-\dfrac{\lambda}{3}\int_{\Omega_c}\! d\Omega\,\big[(R_0+u)^3-R_0^3\big]\\
	+\dfrac{\lambda}{3}\left(\int_{\Omega_c\setminus\Omega_0}\! d\Omega\, -\int_{\Omega_0\setminus\Omega_c}\! d\Omega\right)
	\big[(R_s(\theta))^3-R_0^3\big]\\
	-\lambda\delta V,
	\label{free_en_pheno}
\end{multline}

\noindent with $d\Omega=d\theta d\phi\sin\theta$ and 
\begin{equation}
	s(u,\nabla_a u)
	=(R_0+u)^2\sqrt{1+(\nabla_a u)^2/(R_0+u)^2}, 
\end{equation}

\noindent where
 \begin{equation}
	\nabla_a:=\boldsymbol{e}_{\theta}\partial_{\theta}+\dfrac{\boldsymbol{e}_{\phi}}{\sin\theta}\partial_{\phi}
\end{equation}

\noindent is the dimensionless \textit{a}ngular gradient on the unit sphere. We distinguish two integration domains: 
\begin{equation}
 \Omega_0=\{(\theta,\phi)\in \mathbb{R}^2\; |\; \phi\in[0,2\pi) \;\wedge\; \theta\in[0,\theta_0]\},
\end{equation}

\noindent which corresponds to the reference droplet shape, and $\Omega_c$ in which the actual droplet interface $u(\Omega)$ is defined:
\begin{equation}
 \Omega_c=\{(\theta,\phi)\in \mathbb{R}^2\; |\; \phi\in[0,2\pi) \;\wedge\; \theta\in[0,\theta_c(\phi)]\}.
\end{equation}

\begin{figure}[ht]
	\centering
	\psfragscanon
	\psfrag{u}[c][c][1.3]{$u$}
	\psfrag{r0}[c][c][1.3]{$R_0$}
	\psfrag{h}[c][c][1.3]{$h$}
	\psfrag{f}[c][c][1.3]{$\boldsymbol{f}$}
	\psfrag{o}[c][c][1.3]{$\Delta \Omega$}
	\psfrag{pi}[c][c][1.3]{$\Pi(\Omega)\boldsymbol{e}_{n}$}
	\includegraphics[width=0.45\textwidth]{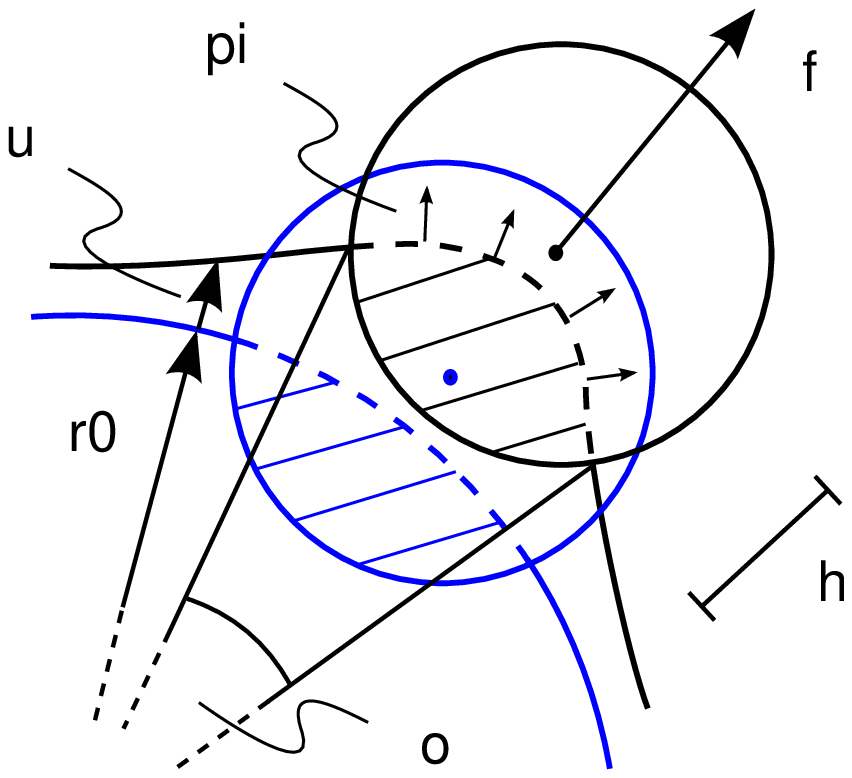}
	\psfragscanoff
	\caption{Schematic cross-sectional representation of the virtual piece of the liquid-gas interface (dashed lines) added inside the particle on which $\Pi(\Omega)$ is defined ($\boldsymbol{e}_{n}$ is a unit vector normal to that interface). The quantity $\delta S_{lg}$ (see main text) is the change in the surface area of that piece relative to the reference configuration ($u=0$), whereas $\delta V$ is the volume change of the corresponding virtual liquid domain inside the particle (hatched regions). The displacement of the center of the particle is denoted as $h$. The radial displacement of the interface relative to the reference configuration ($R_0$) is denoted as $u$.}
	\label{virtual}
\end{figure}

\noindent $\Omega_c$ differs from $\Omega_0$ only by the maximal polar angle $\theta_c=\theta_c(\phi)$, which determines the shape of the \textit{c}ontact line. Equivalently, the shape of the contact line is also described by the deformation $u_c=u_c(\phi)\equiv u(\theta=\theta_c(\phi),\phi)$. In the case of a pinned contact line one has $\Omega_c=\Omega_0$ and $u_c=0$, but in the case of a free contact line in general $\Omega_c\neq\Omega_0$. We note that the parameterization in terms of spherical coordinates remains valid for $\theta_0>\pi/2$ (in this case the center of the droplet lies above the substrate). The first two terms on the rhs of Eq.\ (\ref{free_en_pheno}) emerge from $\gamma[\int_{\Omega_c}\!d\Omega\,s-\int_{\Omega_0}\!d\Omega\,s_{ref}]$ with $s_{ref}=R_0^2$ and thus they represent the changes in the surface energy of the liquid-gas interface. The third term represents the changes in the liquid-substrate surface energy. The fourth term represents the work done by the external pressure $\Pi(\Omega)$ in displacing the interface and the last three terms correspond to the volume conservation. The last but one term corrects the previous one with a liquid volume wedged between the substrate surface and the surface of the cap of the reference sphere inside the domains $\Omega_c\setminus\Omega_0$ and $\Omega_0\setminus\Omega_c$ (in the former case the reference surface is extended into the domain $\Omega_c\setminus\Omega_0$). Accordingly, $R_s(\theta)=R_0\cos\theta_0/\cos\theta$ expresses the distance between points on the \textit{s}ubstrate surface and the origin in terms of spherical coordinates. The last term represents a correction $\delta V$ corresponding to the change of the volume (due to an interface displacement $u$) of the virtual liquid domain added inside the particle (see Fig.\ \ref{virtual}). Comparing the expressions in Eqs.\ (\ref{free_en_pheno}) and (\ref{functional}) yields an implicit definition of the effective pressure field $\Pi$, in the sense that $\Pi(\Omega)$ must be chosen such that the following equation is fulfilled:
\begin{equation}
\dfrac{1}{3}\int_{\Delta\Omega}\! d\Omega\,\Pi(\Omega)\big[(R_0+u)^3-R_0^3\big] = fh +\gamma\delta S_{pl}\cos\theta_p+\gamma \delta S_{lg},
	\label{pressure}
\end{equation}

\noindent where $\delta S_{pl}=S_{pl}-S_{pl,ref}$; $\delta S_{lg}$ is the change of the area of the virtual liquid-gas interface relative to the reference configuration (see Fig.~\ref{virtual}). Equation (\ref{pressure}) does not uniquely determine  $\Pi(\Omega)$ but in the limit of small particles, as will be studied in Subsec.\ III.D, it provides a condition sufficient for calculating the shape of the droplet. 

The shift of the center of mass is given by
\begin{multline}
 	x_{CM}-x_{CM,ref}\,=\\
	=\,\dfrac{1}{V_l}\left(\int_{\Omega_c}\!d\Omega\,\int_{R_s(\theta)}^{R_0+u}\!dr 
	-\int_{\Omega_0}\!d\Omega\,\int_{R_s(\theta)}^{R_0}\!dr\right)\,r^3\sin\theta\cos\phi\,\\
	-\,\delta x \\
	=\,\dfrac{1}{4V_l}\int_{\Omega_c}\!d\Omega\,\big[(R_0+u)^4-R_0^4\big]\sin\theta\cos\phi\,\\
	-\,\dfrac{1}{4V_l}\left(\int_{\Omega_c\setminus\Omega_0}\! d\Omega\, -\int_{\Omega_0\setminus\Omega_c}\! d\Omega\right) \big[(R_s(\theta))^4-R_0^4\big]\sin\theta\cos\phi\,\\
	-\,
	\delta x,
	\label{xcmu}
\end{multline}

\noindent where the second term in the final expression represents the contribution to the shift due to the deformation of the contact line and $\delta x$ is the contribution to the shift of the center of mass of the whole droplet due to the shift of the virtual liquid domain assigned to the inside of the particle. $R_s(\theta)$ has the same meaning as in Eq.\ (\ref{free_en_pheno}).

\section{Perturbation theory for large droplets}

\subsection{Systematic expansion, variation, boundary conditions, and force balance}
It is the aim of this section to base our analytical approach on a systematic expansion in terms of a small parameter $\epsilon$, such that the deformations of the droplet are small, too. We first note that strongly deformed droplets are usually metastable or unstable (see also Subsec.V.A), so that a necessary condition for the deformations being small is the stability of the droplet shape. In the case of a spherical droplet this means that the shift in the internal pressure due to the force $f$ acting on the interface must be much smaller than the Laplace pressure $2\gamma/R_0$. Up to a geometrical factor the shift is of the order of $f/R_0^2$, so that in order to avoid such instabilities one must have
\begin{equation}
	|f| \ll \gamma R_0,
\label{f_small1}
\end{equation} 

\noindent which means that the external force must be much smaller than the maximal capillary force acting on the contact line at the substrate which is of the order of $\gamma R_0$. On the other hand the same reasoning applies to the contact line at the particle implying that the force $f$ should not be stronger than the maximal capillary force acting on the particle which is of the order of $\gamma a$. One can argue that even if the deformation of the interface is large (in a not yet precisely defined sense) near the particle, perturbation theory could still be valid far away from the particle. Thus, for the perturbation theory to be valid, instead of $f \ll \gamma a$ one only needs
\begin{equation}
	|f| \lesssim \gamma a.
\label{f_small2}
\end{equation} 

\noindent Therefore, it is reasonable to choose 
\begin{equation}
	\epsilon = \dfrac{|f|}{\gamma R_0}
\label{epsilon}
\end{equation}

\noindent as a small dimensionless parameter, assuming that the condition in Eq.\ (\ref{f_small2}) is fulfilled, too. We note that for large droplets, i.e., for 
\begin{equation}
	a\ll R_0,
\label{allR0}
\end{equation}

\noindent the condition  $\epsilon\ll1$ is automatically satisfied.

In the next step, we expand $u$, $\lambda$, $\Pi$, $f_{CM}$, and, for later purposes, $h$ in terms of the small parameter $\epsilon$:
\begin{align}
	&u(\Omega)=R_0\big(\epsilon v(\Omega) + o(\epsilon) \big),\label{expu}\\
	&\lambda=\frac{\gamma}{R_0}\big(2+\epsilon\mu + o(\epsilon) \big),\\
	&\Pi(\Omega)=\frac{\gamma}{R_0}\big(\epsilon\pi(\Omega)+ o(\epsilon) \big),\label{exppi}\\	
	&f_{CM}=\gamma R_0\big(\epsilon Q_{CM}+ o(\epsilon) \big),\label{expfcm}\\
	&h=R_0\big(\epsilon\tilde{h} + o(\epsilon) \big),\label{exph}
\end{align}

\noindent where $\lambda$ has been expanded around the Laplace pressure $2\gamma/R_0$ of a spherical droplet; $v$, $\mu$, $\pi$, $Q_{CM}$, and $\tilde{h}$ are dimensionless. 

In the following we shall keep the radius $R_0$ of the droplet fixed and therefore it is convenient to divide the energy by $\gamma R_0^2$. In zeroth order in $\epsilon$ the droplet is undeformed ($u\equiv 0$) and the free energy functional  $\mathcal{F}^{(0)}$ as defined in Eq.\ (\ref{free_en_pheno}) equals zero.  Similarly, the free energy $\mathcal{F}^{(1)}$ in first order in $\epsilon$ also vanishes, which reflects the fact that we have chosen the equilibrium configuration as the reference one (in equilibrium, by definition, perturbations do not give rise to linear contributions to the free energy). Therefore the leading contribution is of second order in $\epsilon$ and the free energy functional reads 
%\begin{equation}
\begin{multline}
	\dfrac{\mathcal{F}}{\gamma R_0^2}= \epsilon^2 \int_{\Omega_0}\! d\Omega\, \left[\frac{1}{2}(\nabla_a v)^2 - v^2 - \big(\pi(\Omega)+\mu\big)v\right]\\
	\,-\,\dfrac{\epsilon^2 }{2}\cos\theta_0\int_0^{2\pi}\!d\phi\, (v|_{\theta_0})^2 + O(\epsilon^3),
	\label{small_def2}
\end{multline}
%\end{equation}

\noindent where $\pi(\Omega\notin\Delta\Omega)=0$ and the boundary term (second term) has been obtained by expanding up to second order in $\epsilon$ those terms in the free energy functional in Eq.\ (\ref{free_en_pheno}), which depend on the deformations of the contact line. (For $\theta_c(\phi)<\theta_0$, the deformation $v(\theta)$ is smoothly extended to $\theta=\theta_0$, so that in Eq.\ (\ref{small_def2}) $v(\theta)$ is defined within the whole angular domain $\Omega_0$; this is justified because the corresponding difference in the free energy is of higher order $O(\epsilon^3)$.) The variation of $\mathcal{F}$ renders
\begin{multline}
	\dfrac{1}{\gamma R_0^2}\big(\mathcal{F}[v+\delta v]-\mathcal{F}[v]\big)=\\
	=\epsilon^2\int_{\Omega_0}\!d\Omega\,[-\nabla_a^2v - 2v - \pi(\Omega) - \mu]\delta v\\
	+\epsilon^2\int_{\Omega_0}\,d\Omega\,\nabla_a\cdot(\delta v\nabla_a v)\\
	-\,\epsilon^2 \cos\theta_0\int_0^{2\pi}\!d\phi\, v|_{\theta_0}\delta v|_{\theta_0} + O((\delta v)^2,\epsilon^3),
	\label{variation}
\end{multline}

\noindent whereas the variation of the additional term responsible for fixing the center of mass (Eq.\ (\ref{xcmu})) yields
\begin{multline}
%\begin{equation}
	-\dfrac{f_{CM}}{\gamma R_0^2}(x_{CM}-x_{CM,ref})|^{v+\delta v}_v\, =\\
	=\, -\epsilon^2\dfrac{3Q_{CM}}{4\pi f_0(\theta_0)}\int_{\Omega_0}\!d\Omega\,\delta v\sin\theta\cos\phi + O((\delta v)^2,\epsilon^3),
\label{xcm_var}
%\end{equation}
\end{multline}

\noindent where we have neglected the second term in Eq.\ (\ref{xcmu}), which is of the order $O(\epsilon^2)$ and thus in Eq.\ (\ref{xcm_var}) it contributes only to the order $O(\epsilon^3)$, and $\delta x$ which contributes only to the order $O(\epsilon^2)\times O(a/R_0)^3$. The function $f_0(\theta_0)$ expresses the volume of a unit spherical cap characterized by the polar angle $\theta_0$ as a fraction of the volume $4\pi/3$ of a unit sphere:
\begin{equation}
 	f_0(x):=(2+\cos x)\sin^4(x/2).
	\label{f0}
\end{equation}

\noindent The expressions in Eqs.\ (\ref{variation}) and (\ref{xcm_var}) lead to the Euler-Lagrange equation for a droplet with a fixed center of mass in the form
\begin{equation}
	-(\nabla_a^2+2)v(\Omega)=\pi(\Omega) + \pi_{CM}(\Omega)  + \mu,
	\label{helmholtz}
\end{equation}

\noindent where 
\begin{equation}
	\pi_{CM}(\Omega)=\dfrac{3Q_{CM}}{4\pi f_0(\theta_0)}\sin\theta\cos\phi
	\label{piCM}
\end{equation}

\noindent is an effective pressure fixing the center of mass of the droplet. In the case that the center of mass is not fixed, the corresponding Euler-Lagrange equation has the same form as in Eq.\ (\ref{helmholtz}) but with $\pi_{CM}=0$. Applying Gau{\ss}' theorem (on a unit sphere) to the second term in Eq.\ (\ref{variation}) gives the \textit{b}oundary \textit{c}ontribution
\begin{multline}
%\begin{equation}
	\dfrac{1}{\gamma R_0^2}\big(\mathcal{F}[v+\delta v]-\mathcal{F}[v]\big)_{bc}=\\
	=\epsilon^2\int_0^{2\pi}\!d\phi\, (\sin\theta_0\partial_{\theta} v|_{\theta_0}-\cos\theta_0v|_{\theta_0})\delta v|_{\theta_0}, \label{Fbc}
%\end{equation}
\end{multline}

\noindent vanishing if $\delta v|_{\theta_0}=0$, which corresponds to a pinned contact line (model $B$). If the contact line is free (model $A$) one has $\delta v|_{\theta_0}\neq0$ and, instead, the expression in brackets of the integrand in Eq.\ (\ref{Fbc}) must vanish. These two different conditions correspond to Dirichlet $(B)$ and Robin $(A)$ boundary conditions, respectively:
\begin{align}
 \, & A\,(\text{free}): & \sin\theta_0\partial_{\theta}v|_{\theta_0}-\cos\theta_0v|_{\theta_0} & = 0, &\  &\ \label{Rob}\\
 \, & B\,(\text{pinned}): & v|_{\theta_0} & = 0. &\  &\ \label{Dir}
\end{align}

\noindent Equation (\ref{Rob}) is equivalent to the condition that the contact angle is equal to the Young angle, which can be seen by the following reasoning. In the limit of small deformations the normal $\boldsymbol{e}_n$ to the droplet surface in terms of spherical coordinates can be expressed as $\boldsymbol{e}_n=\boldsymbol{e}_r-\epsilon\nabla_a v+O(\epsilon^2)$, so that the contact angle $\tilde{\theta}(\phi)$ at the substrate is given by
%\begin{equation}
\begin{multline} \cos\tilde{\theta}(\phi)=\boldsymbol{e}_z\cdot\boldsymbol{e}_n|_{\theta_c}=\cos\theta_c+\epsilon\sin\theta_c\partial_{\theta}v|_{\theta_c}+O(\epsilon^2)=\\
=\cos\theta_0+\epsilon[\sin\theta_0\partial_{\theta}v|_{\theta_0} -v|_{\theta_0}\cos\theta_0]+O(\epsilon^2),
\label{tildetheta0}
\end{multline}
%\end{equation}

\noindent where in the third equality we used $\cos\theta_c=\cos\theta_0(1-\epsilon v|_{\theta_0})+O(\epsilon^2)$, which follows from the analysis of small perturbations of a spherical cap. Thus, from Eqs.\ (\ref{Rob}) and (\ref{tildetheta0}) we recover Young's law in the form $\tilde{\theta}(\phi)=\theta_0$. We note that whereas the condition in Eq.\ (\ref{Dir}) corresponds to the shape of the droplet at the actual contact line, because for a pinned contact line $\theta_c(\phi)=\theta_0$, the condition in Eq.\ (\ref{Rob}) does not apply directly at the contact line $\partial \Omega_c$ but at the boundary $\partial\Omega_0$ of the reference integration domain. The actual shape of the contact line in this case can be obtained by a linear extrapolation of the interface profile from $\theta_0$ to $\theta_c$. 

The volume constraint in Eq.\ (\ref{vol_constr}) in first order in $\epsilon$ reads
\begin{equation}
	 \int_{\Omega_0}\!  d\Omega\, v(\Omega) = 0,
	\label{volume2}
\end{equation}

\noindent whereas in Eq.\ (\ref{xcm_constr}) the constraint of fixed center of mass, due to Eqs.\ (\ref{xcmu}) and (\ref{piCM}), can be written, up to second order in $\epsilon$, as
\begin{equation}
	 \int_{\Omega_0}\! d\Omega\, \pi_{CM}(\Omega)v(\Omega) = 0.
	\label{xcmu2}
\end{equation}

\noindent In Eqs.\ (\ref{volume2})  and (\ref{xcmu2}) we have neglected contributions from $\delta V$ (Eq.\ (\ref{free_en_pheno})) and $\delta x$ (Eq.\ (\ref{xcmu})), which are both of the order $O(a/R_0)^3$. 

In the following our goal is to express the pressure shift $\mu$ and the balance of forces acting on the droplet in terms of $\pi$ and $\theta_0$. Integrating both sides of Eq.\ (\ref{helmholtz}) over $\Omega_0$, with $\int_{\Omega_0}\!d\Omega\,=2\pi(1-\cos\theta_0)$, and applying Gau{\ss}' theorem and subsequently the condition of constant volume (Eq.\ (\ref{volume2})) yields 
\begin{equation} \mu=-\dfrac{1}{2\pi(1-\cos\theta_0)}\left[\int_{\Omega_0}\!d\Omega\,\pi(\Omega)+\sin\theta_0\int_0^{2\pi}\!d\phi\,\partial_{\theta}v|_{\theta_0} \right],
\label{muD}
\end{equation}

\noindent where we have used the fact that $\int_{\Omega_0}\!d\Omega\,\pi_{CM}=0$. Equation (\ref{muD}) can be interpreted as a hydrostatic balance in that the internal pressure (lhs) is equal to the external pressure exerted by the forces acting at the droplet (rhs), which in particular depend on the boundary conditions. 

The governing equation (\ref{helmholtz}) can also be used to derive the force balance on the droplet. First, multiplying both sides by the radial vector $\boldsymbol{e}_r(\Omega)$ and integrating over $\Omega_0$ we obtain 
\begin{multline}
%\begin{equation}
	-\int_{\Omega_0}\!d\Omega\, \boldsymbol{e}_r(\Omega)(\nabla_a^2+2)v =\\
	= \int_{\Omega_0}\!d\Omega\,\pi(\Omega)\boldsymbol{e}_r(\Omega)\,+\,Q_{CM}\boldsymbol{e}_x\,+\, \mu\pi\sin^2\theta_0\boldsymbol{e}_z.
	\label{intD0}
%\end{equation}
\end{multline}

\noindent The first term on the rhs of Eq.\ (\ref{intD0}) represents, up to first order in $\epsilon$, the total external force acting on the droplet. The surface integral on the lhs, after integrating by parts, using the fact that $(\nabla_a^2+2)\boldsymbol{e}_r(\Omega)=0$, which follows from the identity $(\nabla_a^2+2)Y_{1m}=0$, and finally applying Gau{\ss}' theorem can be transformed into a line integral,
\begin{multline}
%\begin{equation}
	-\int_{\Omega_0}\!d\Omega\, \boldsymbol{e}_r(\Omega)(\nabla_a^2+2)v =\\
	= -\sin\theta_0\int_{0}^{2\pi}\!d\phi\, (\boldsymbol{e}_r\partial_{\theta} v - v\partial_{\theta}\boldsymbol{e}_r)|_{\theta_0}.
\label{int}
%\end{equation}
\end{multline}

\noindent  Using $\partial_{\theta}\boldsymbol{e}_r=\boldsymbol{e}_{\theta}$ and taking the limit $a/R_0\rightarrow0$ Eq.\ (\ref{intD0}) can be expressed in terms of Cartesian coordinates: 
\begin{align}
\begin{split}
 x:   -\sin\theta_0\int_0^{2\pi}\!d\phi\,\cos\phi \big(\sin\theta_0\partial_{\theta}v|_{\theta_0} -v|_{\theta_0}\cos\theta_0 \big) = \\
 =\int_{\Omega_0}\!d\Omega\,\pi(\Omega)\sin\theta\cos\phi+Q_{CM},	\label{fbal_x}
\end{split}\\
\begin{split}
 y:   -\sin\theta_0\int_0^{2\pi}\!d\phi\,\sin\phi \big(\sin\theta_0\partial_{\theta}v|_{\theta_0} -v|_{\theta_0}\cos\theta_0 \big) = \\ 
 =0,  \label{fbal_y}
\end{split}\\
\begin{split}
 z:  -\sin\theta_0\int_0^{2\pi}\!d\phi\, \big(\cos\theta_0\partial_{\theta}v|_{\theta_0} +v|_{\theta_0}\sin\theta_0 \big) = \\
 =\int_{\Omega_0}\!d\Omega\,\pi(\Omega)\cos\theta+\mu\pi\sin^2\theta_0.  \label{fbal_z}
\end{split}	
\end{align}

\noindent In the case of a free contact line (see Eq.\ (\ref{Rob})) the left hand sides of Eqs.\ (\ref{fbal_x}) and (\ref{fbal_y}) vanish and the force balance in the $x$-direction reduces to
\begin{equation}
 -Q_{CM}=\int_{\Omega_0}\!d\Omega\,\pi(\Omega)\sin\theta\cos\phi.
\label{qcm}
\end{equation}

\noindent If $\pi(\Omega)$ corresponds to a pointlike particle (see, c.f., Eq.\ (\ref{pressure3})), so that $\pi(\Omega)=q\delta(\Omega,\Omega_1)=q\delta(\theta-\theta_1)\delta(\phi-\phi_1)/\sin\theta$, where $\Omega_1=(\theta_1=\alpha,\phi_1=0)$ is the direction along which the external force pulls ($q=+1$) or pushes ($q=-1$) the particle,  due to $f_{CM}=\epsilon \gamma R_0 Q_{CM}=|f|Q_{CM}$  (see Eqs.\ (\ref{epsilon}) and (\ref{expfcm})) Eq.\ (\ref{qcm}) leads to $f_{CM}=-f\sin\alpha$. Accordingly, the Lagrange multiplier $f_{CM}$ (Eq.\ (\ref{minimizeN})) can indeed be interpreted as a force counterbalancing the $x$-component of the external force, equal to $f\sin\alpha$, and thus fixing the center of mass of the droplet. In the same limiting case the first term on the rhs of Eq.\ (\ref{intD0}) reduces to $\int_{\Omega_0}\!d\Omega\,\pi(\Omega)\boldsymbol{e}_r(\Omega)=q\boldsymbol{e}_r(\Omega_1)$ corresponding to the external force on the droplet. In the case of a pinned contact line and with a free center of mass (i.e., $f_{CM}=0$ so that $Q_{CM}=0$) the force balance in the $x$-direction (Eq.\ (\ref{fbal_x})) reads
\begin{equation} -\sin^2\theta_0\int_0^{2\pi}\!d\phi\,\cos\phi\partial_{\theta}v|_{\theta_0}=\int_{\Omega_0}\!d\Omega\,\pi(\Omega)\sin\theta\cos\phi.
\label{fbal_xB}
\end{equation}

\noindent Thus, in this case, the $x$-component of the external force (rhs) is counterbalanced by the capillary force due to the deformation of the droplet at the contact line (lhs). 

For model $A$ the lhs of Eq.\ (\ref{fbal_y}) vanishes due to Eq.\ (\ref{Rob}), but it also vanishes for model $B$ because $\partial_{\theta}v|_{\theta_0}$ and $v|_{\theta_0}$ are symmetric and $\sin\phi$ is antisymmetric with respect to the $xz$-plane. 

The difference between these two models is that for model $A$ the local contact angle is the Young angle (see Eqs.\ (\ref{tildetheta0}) and (\ref{Rob})), which guarantees that each piece of the contact line is in mechanical equilibrium, whereas in the case of model $B$ this is not the case and the total capillary force on the contact line counterbalances the $x$-component of the external force (Eq.\ (\ref{fbal_xB})). 

The boundary condition in Eq.\ (\ref{Rob}) allows one to express $\partial_{\theta}v|_{\theta_0}$ in terms of $v|_{\theta_0}$. Inserting this into Eqs.\ (\ref{muD}) and (\ref{fbal_z}) renders two equations for $\mu$ and $\int_0^{2\pi}\!d\phi\,v|_{\theta_0}$, which can be solved to yield
\begin{align}
	\, & A: & \mu & = -\dfrac{1}{4\pi f_0(\theta_0)}\int_{\Omega_0}\!d\Omega\,\pi(\Omega)(1-\cos\theta_0\cos\theta), &\  &\ \label{musigmaA}\\
	\, & B: & \mu & = -\dfrac{1}{\pi(1-\cos\theta_0)^2}\int_{\Omega_0}\!d\Omega\,\pi(\Omega)(\cos\theta-\cos\theta_0). &\  &\ \label{musigmaB}
\end{align}

\noindent Equation (\ref{musigmaB}) follows from inserting the boundary condition in Eq.\ (\ref{Dir}) into Eq.\ (\ref{fbal_z}); this renders $\int_0^{2\pi}\!d\phi\,\partial_{\theta}v|_{\theta_0}$ in terms of $\mu$, which can be inserted into Eq.\ (\ref{muD}) yielding an equation for $\mu$. In the case of a pointlike particle (see, c.f.,  Eq.\ (\ref{pressure3})), Eqs.\ (\ref{musigmaA}) and (\ref{musigmaB}) reduce to $\mu=-q(1-\cos\theta_0\cos\alpha)/(4\pi f_0(\theta_0))$ and $\mu=-q(\cos\alpha-\cos\theta_0)/(\pi(1-\cos\theta_0)^2)$, respectively, so that due to $\alpha<\theta_0$ (see Fig.\ \ref{sketch}$(a)$) the internal uniform pressure shift $\mu$ has always the sign $-q$, i.e., the opposite one to $f$. This means that the pressure change counteracts the action of the external force, which can be interpreted as the realization of Le Chatelier's principle for this particular system.

\subsection{Green's functions}
In this subsection we present a formal solution of Eq.\ (\ref{helmholtz}) by means of Green's functions $G_{\sigma}(\Omega,\Omega')$ which are the radial deformations $v(\Omega)$ due to a pointlike perturbation  $\pi(\Omega)=\delta(\Omega,\Omega')=\delta(\theta-\theta')\delta(\phi-\phi')/\sin\theta$ acting in the direction $\Omega'=(\theta',\phi')$. 
In this respect we point out that in the case of arbitrary $\Omega'$ (i.e., a direction pointing out of the $xz$-plane) the term $-f_{CM}(x_{CM}-x_{CM,ref})$ in the free energy functional in Eq.\ (\ref{minimizeN}) has to be replaced by $-f_{CM,x}(x_{CM}-x_{CM,ref})-f_{CM,y}(y_{CM}-y_{CM,ref})$ where $f_{CM,x}$ and $f_{CM,y}$ are the corresponding Lagrange multipliers. Accordingly, in analogy to Eq.\ (\ref{piCM}), in this case one has $\pi_{CM}(\Omega)=3\sin\theta[Q_{CM,x}\cos\phi+Q_{CM,y}\sin\phi]/(4\pi f_0(\theta_0))$ with $Q_{CM,x}:=f_{CM,x}/|f|\equiv Q_{CM}$ given by Eq.\ (\ref{qcm}) and $Q_{CM,y}:=f_{CM,y}/|f| =-\int_{\Omega_0}\!d\Omega\,\pi(\Omega)\sin\theta\sin\phi$. This means that Green's functions fulfill Eq.\ (\ref{helmholtz}) with $\pi_{CM}(\Omega)=3\sin\theta\sin\theta'\cos(\phi-\phi')/(4\pi f_0(\theta_0))$ for model $A$, $\pi_{CM}=0$ for model $B$, and $\mu$ given by Eqs.\ (\ref{musigmaA}) and (\ref{musigmaB}). Accordingly Green's functions fulfill 
\begin{multline}
%\begin{equation}
 -(\nabla_a^2+2)G_A(\Omega,\Omega')=\delta(\Omega,\Omega')\\
  - \dfrac{3}{4\pi f_0(\theta_0)}\sin\theta'\sin\theta\cos(\phi-\phi')
  -\dfrac{1-\cos\theta_0\cos\theta'}{4\pi f_0(\theta_0)}  
        \label{govGN}
%\end{equation}
\end{multline}

\noindent for model $A$ and
\begin{equation}
-(\nabla_a^2+2)G_B(\Omega,\Omega')=\delta(\Omega,\Omega') - \dfrac{\cos\theta'-\cos\theta_0}{\pi(1-\cos\theta_0)^2}
	\label{govGD}
\end{equation}

\noindent for model $B$, where in both cases $\Omega,\Omega'\in\Omega_0$. We note that for arbitrary $\theta_0$ the Green's functions are not symmetric, i.e., $G_{\sigma}(\Omega,\Omega')\neq G_{\sigma}(\Omega',\Omega)$. This means that the deformation at $\Omega$ due to a perturbation at $\Omega'$ in general differs from the deformation at $\Omega'$ due to a perturbation at $\Omega$. This can be understood by the fact that the shift $\mu$ of the internal pressure depends on where the perturbation is applied. The only exception is the case $\theta_0=\pi/2$ with a free contact line, for which the Green's function $G_A$ is fully symmetric. In this case $\cos\theta_0=0$ so that the last term on the rhs of Eq.\ (\ref{govGN}) is a constant, whereas the last but one term is explicitly symmetric.

The deformation of the droplet for an arbitrary pressure field $\pi(\Omega)$ can be expressed in terms of the Green's functions:
\begin{equation}
 v_{\sigma}(\Omega)=\int_{\Omega_0}\!d\Omega'\,G_{\sigma}(\Omega,\Omega')\pi(\Omega').
\label{ND-solution}
\end{equation}

\noindent By using Eqs.\ (\ref{govGN}) and (\ref{govGD}) one can check that this expression for $v_{\sigma}$ fulfills the Young-Laplace equation (\ref{helmholtz}). For $\pi(\Omega)=\delta(\Omega,\Omega')$ the volume constraint in Eq.\ (\ref{volume2}) yields 
\begin{equation}
 \int_{\Omega_0}\!d\Omega\,G_{\sigma}(\Omega,\Omega')=0,
	\label{volG}
\end{equation}

\noindent whereas the boundary conditions in Eqs.\ (\ref{Rob}) and (\ref{Dir}) can be expressed as
\begin{align} \sin\theta_0\partial_{\theta}G_A(\Omega,\Omega')\vert_{\Omega\in\partial\Omega_0}-\cos\theta_0G_A(\Omega,\Omega')\vert_{\Omega\in\partial\Omega_0} &= 0,\label{boundary_N}\\
 G_B(\Omega,\Omega')\vert_{\Omega\in\partial \Omega_0} &= 0.\label{boundary_D}
\end{align}

\subsection{Free energy}
It can be shown that for the boundary conditions of either a pinned or a free contact line the total boundary contribution to the free energy vanishes (see Appendix A). In analogy to electrostatics (see, for example, p.\ 46 in Ref.~\cite{Jackson}) we obtain the following expression (derived in Appendix A) for the free energy of a sessile droplet subjected to an arbitrary external pressure $\pi(\Omega)$:
\begin{multline}
%\begin{equation}
	F_{\sigma} =- \frac{f^2}{2\gamma}\int_{\Omega_0}\! d\Omega\,\pi(\Omega)v(\Omega) =\\
	= - \frac{f^2}{2\gamma}\int_{\Omega_0} \!d\Omega\, \int_{\Omega_0} \!d\Omega'\, \pi(\Omega)G_{\sigma}(\Omega,\Omega')\pi(\Omega').
\label{F_ND}
%\end{equation}
\end{multline}

\noindent where for the second equality we have used the general form of the solution given in Eq.\ (\ref{ND-solution}).

\subsection{Limit of small particles}
Taking the limit $a\rightarrow0$ allows one to provide a relation between the unknown pressure field $\pi$ and the system parameters introduced in Sec.\ II. In this context, Eq.\ (\ref{pressure}) provides a constraint on $\pi$ expressed in terms of an unknown deformation field, encoded by the quantities $h$, $\delta S_{pl}$, and $\delta S_{lg}$. However, this equation simplifies considerably in the limit of small particles. In order to see this, we consider Eqs.\ (\ref{govGN}) and (\ref{govGD}) in the limit $\Omega\rightarrow\Omega'$ {in which} the spherical reference interface becomes locally flat in the neighborhood of $\Omega$ and $\Omega'$. First, we rotate the coordinate frame such that the $z$-axis points in an arbitrarily chosen direction in the neighborhood of $\Omega$ and $\Omega'$. (In the following we refer to this new axis as the $\tilde{z}$-axis.) Next, we project {the points} $\Omega$ and $\Omega'$ {on the unit sphere} along the $\tilde{z}$-axis onto the plane tangent to the reference spherical interface at the point of intersection of this interface with the $\tilde{z}$-axis. The mapping $(\tilde{\theta},\tilde{\phi})\mapsto(\tilde{\rho},\tilde{\phi})$, where $(\tilde{\theta},\tilde{\phi})$ denote the spherical coordinates associated with the $\tilde{z}$-axis and $(\tilde{\rho},\tilde{\phi})$ are polar coordinates in the tangent plane, is given by 
\begin{equation}
 \tilde{\rho}=R_0\sin\tilde{\theta}.
\label{rhotheta}
\end{equation}

\noindent {In order to keep the notation simple, in the following} we skip the tilde so that $(\tilde{\rho},\tilde{\phi})\equiv(\rho,\phi)$. Under the transformation of coordinates the Laplace operator transforms according to the chain rule whereas the delta function transforms as: $\delta(\phi-\phi')\,\delta(\theta-\theta')/\sin\theta = R_0\,\delta(\phi-\phi')\,\delta(\theta(\rho)-\theta'(\rho'))/\rho = R_0\,|d\rho/d\theta|\,\delta(\phi-\phi')\,\delta(\rho-\rho')/\rho$. Accordingly, Eqs.\ (\ref{govGN}) and (\ref{govGD}) can be rewritten in terms of a single equation (for details see Ref.~\cite{GuzowskiPhd}):
\begin{multline}
%\begin{equation}
	-(R_0^2\,\nabla_{\parallel}^2+2-\partial_{\rho}\,\rho^2\,\partial_{\rho})G_{\sigma}(\boldsymbol{x},\boldsymbol{x}')=\\
	=R_0\sqrt{R_0^2-\rho^2}\,\delta(\boldsymbol{x}-\boldsymbol{x}')+\Delta_{\sigma}(\boldsymbol{x},\boldsymbol{x}'),
\label{Gparallel}
%\end{equation}
\end{multline}

\noindent where {$\boldsymbol{x}$ denotes the Cartesian coordinates within the tangent plane,} $\nabla_{\parallel}^2$ is the Laplace operator on the tangent plane, and $G_{\sigma}(\boldsymbol{x},\boldsymbol{x}'):=G_{\sigma}(\Omega(\boldsymbol{x}),\Omega'(\boldsymbol{x}'))$; the function $\Delta_{\sigma}(\boldsymbol{x},\boldsymbol{x}'):=\Delta_{\sigma}(\Omega(\boldsymbol{x}),\Omega'(\boldsymbol{x}'))$ {stands for all non-singular} terms on the right hand sides of Eqs.\ (\ref{govGN}) and (\ref{govGD}), where $\boldsymbol{x}(\Omega)=(x=R_0\sin\tilde{\theta}\cos\tilde{\phi},y=R_0\sin\tilde{\theta}\sin\tilde{\phi})$ describes the rotation $\Omega\mapsto\tilde{\Omega}$, which might be expressed in terms of the corresponding Euler angles, with the subsequent projection onto the tangent plane $\tilde{\Omega}\mapsto\boldsymbol{\tilde{x}}\equiv\boldsymbol{x}$ (see also Eq.\ (\ref{rhotheta})). In the limit $R_0\rightarrow \infty$ Eq.\ (\ref{Gparallel}) {reduces} to the usual two-dimensional Green's equation
\begin{equation}
	-\nabla_{\parallel}^2G_{\sigma}(\boldsymbol{x},\boldsymbol{x}')
	=\delta(\boldsymbol{x}-\boldsymbol{x}'),
\label{Gparallel2}
\end{equation}

\noindent with the solution  $G_{\sigma}(\boldsymbol{x},\boldsymbol{x}')\equiv G(\boldsymbol{x},\boldsymbol{x}')=G(|\boldsymbol{x}-\boldsymbol{x}'|)=-(1/(2\pi)) \ln(|\boldsymbol{x}-\boldsymbol{x}'|)$ which depends neither on $\sigma$ nor separately on $\boldsymbol{x}'$, or equivalently $\Omega'$, i.e., where on the droplet the perturbation is applied.

{Therefore} the deformation $v_{\sigma}$ diverges logarithmically in the neighborhood of a pointlike perturbation at $\Omega'$, independently of the boundary conditions far away from $\Omega'$. By taking $\pi(\Omega)=\delta(\Omega,\Omega')$ in Eq.\ (\ref{ND-solution}) one obtains 
\begin{equation}
 v_{\sigma}(\Omega)=G_{\sigma}(\Omega,\Omega')\xrightarrow[\Omega\rightarrow\Omega']{}-\dfrac{1}{2\pi}\ln(\bar{\theta}).
\end{equation}

\noindent where $\bar{\theta}$ is the angle between the unit vectors pointing into the directions $\Omega$ and $\Omega'$. Thus, with the radius $a$ of the particle acting as a natural cutoff, in leading order in $a/R_0$ the dimensionless displacement $\tilde{h}$ of the particle (see Eq.\ (\ref{exph}))  reads
\begin{equation}
 \tilde{h}=\dfrac{q}{2\pi}\ln\left(\dfrac{R_0}{a}\right)+{O(1)},
\label{tildeh}
\end{equation}

\noindent where $q=f/|f|=\text{sgn}(f)$. Accordingly Eq.\ (\ref{pressure}) in leading order in $\epsilon$ can be written as (see Eqs.\ (\ref{expu}), (\ref{exppi}), (\ref{exph}), and (\ref{epsilon}))
\begin{equation}
 	\int_{\Omega_0}\! d\Omega\,\pi(\Omega)v(\Omega)\, = \dfrac{1}{2\pi}\ln\left(\dfrac{R_0}{a}\right)+ O(1). %O(\gamma^2a^2/f^2),
	\label{pressure2}
\end{equation}

\noindent The correction term in Eq.\ (\ref{pressure2}), which stems from the last two terms on the rhs of Eq.\ (\ref{pressure}), is proportional to $\gamma^2a^2/f^2$ which, in agreement with Eq.\ (\ref{f_small2}), is of order 1 relative to the leading term which diverges $\sim\ln(R_0/a)$. Moreover, assuming that the virtual piece of the interface defined at $\Delta \Omega$ is bounded to lie inside the particle, one has $u(\Omega)|_{\Delta \Omega}=h+O(a/R_0)$ (see Fig. \ref{virtual}) so that $v(\Omega)|_{\Delta \Omega}=\tilde{h}+O(a/R_0)$. According to Eqs.\ (\ref{tildeh}) and (\ref{pressure2}) this leads to
\begin{equation}
 	\int_{\Omega_0}\! d\Omega\,\pi(\Omega)\, = \, q + O(1/\ln(R_0/a)).
	\label{pressure3}
\end{equation}

\noindent Equations (\ref{pressure2}) and (\ref{pressure3}) are satisfied by $\pi(\Omega)=q\,\delta(\Omega,\Omega_1)$ with $\Omega_1=(\theta=\alpha,\phi=0)$ (see Fig.\ \ref{sketch}). Thus, in leading order in $a/R_0$ the integrated amplitude of the effective pressure is independent of the particle radius. In this limit the droplet radius $R_0$ is the only remaining length scale.

According to the above considerations, for $\Omega\rightarrow\Omega'$ Green's functions $G_{\sigma}(\Omega,\Omega')$ in Eq.\ (\ref{F_ND}) contain a logarithmically diverging part $G(\Omega,\Omega')\sim\ln\bar{\theta}$. Therefore it is useful to consider this singular contribution separately by writing $G_{\sigma}=G+G_{\sigma,reg}$, such that the regular part $G_{\sigma,reg}=G_{\sigma}-G$ does not diverge for $\Omega\rightarrow\Omega'$. (As described in the subsequent section in special cases $G$ can be identified as Green's function for a free droplet (see, c.f., Eq.\ (\ref{greens_function})).) 

With $\pi(\Omega)=q\,\delta(\Omega,\Omega_1)$ this leads to the following expressions for the free energy {(Eq.\ (\ref{F_ND}))}:
\begin{equation}
 	F_{\sigma0} = 
	-\dfrac{f^2}{2\gamma}\big[G(\bar{\theta}=a/R_0)+G_{\sigma,reg}(0,0)+O(1)\big]
	%-\dfrac{f^2}{2\gamma}G_{\sigma}(0,0)|_{reg}+O(a^2).
\label{Fs0}
\end{equation}

\noindent and
\begin{equation}
 	\Delta F_{\sigma} = 
	-\dfrac{f^2}{2\gamma}\big[G_{\sigma,reg}(\Omega_1,\Omega_1)-G_{\sigma,reg}(0,0)+O(a/R_0)\big]
\label{DFs}
\end{equation}

\noindent where {$G(\Omega=\Omega_1,\Omega'=\Omega_1)$} for $\Omega_1=0$ (the apex position) has been regularized as $G(\bar{\theta}=a/R_0)$. The leading term of the free energy $F_{\sigma0}$ stems from the singular part of the kernel in {(Eq.\ (\ref{F_ND}))} and thus, in analogy to electrostatics, it can be interpreted as a \textit{self-energy} of the particle. However, unlike the self-energy of a {pointlike charge} in electrostatics, this quantity does not diverge because the singularity is suppressed by the prefactor $f^2$ Eq.\ (\ref{Fs0}) the upper bound of which, as already noted, is proportional to $\gamma^2a^2$ (Eq.\ (\ref{f_small2})). Thus for $a\rightarrow0$ with $f/(\gamma a)$ kept constant the self-energy $F_{\sigma0}$ vanishes as $a^2\ln(R_0/a)$.  In the following section we shall rather focus on the remaining part of the free energy $\Delta F_{\sigma}$ which determines the angular free energy landscape of the particle.

Comparing the first equality in Eq.\ (\ref{F_ND}) with Eq.\ (\ref{pressure2}) one can see that a correction of order $O(1)$ due to the surface energy of the particle (see Eq.\ (\ref{pressure})) also contributes to the free energy $F_{\sigma}$. As indicated in Eqs.\ (\ref{Fs0}) and (\ref{DFs}), {for} $\gamma F_{\sigma0}/f^2$ this correction contributes to the order $O(1)$ but {for} $\gamma \Delta F_{\sigma}/f^2$ it contributes only to the order $O(a/R_0)$ or higher. This is the case because the changes of the surface energy of the particle with respect to $\alpha$ occur only due to the presence of the substrate (the free energy involving a free droplet would not depend on $\alpha$) which is located at a distance $O(R_0)$ from the particle and thus gives rise to the above factor $O(a/R_0)$.

%%%%%%%%%%%%%%%%%%%%%%%%%%%%%%%%%%%%%%%%%%%%%%%%%%%%%%%%%%%%%%%%%%%%%%%%%%%%%%%%%%%%%
%%%%%%%%%%%%%%%%%%%%%%%%%%%%%%%%%%%%%%%%%%%%%%%%%%%%%%%%%%%%%%%%%%%%%%%%%%%%%%%%%%%%%

\section{Method of images for $\theta_0=\pi/2$}

For the special case $\theta_0=\pi/2$ one can construct the Green's functions $G_{\sigma}$ by using the method of images. First, in the reference configuration, the substrate is replaced by a virtual mirror image (with respect to the surface plane of the substrate) of the actual reference droplet. We obtain a \textit{full} and free sphere composed of the \textit{actual} upper hemisphere and the \textit{virtual} lower hemisphere. Next, the deformation of the \textit{actual} droplet subjected to a pointlike force can be constructed as the upper part of the \textit{full} droplet subjected to this pointlike force plus additional virtual pointlike forces, subsequently called images, placed at the \textit{virtual} lower hemisphere. The distribution of these images has to be chosen such that the following conditions are satisfied:
%\vspace*{0.5cm}

\begin{itemize}
 \item[$(i)$] the shape of the droplet, given at a direction $\Omega$ by $G_{\sigma}(\Omega,\Omega')$, obeys the boundary conditions in Eqs.\  (\ref{boundary_N}) or (\ref{boundary_D}), depending on the model;

 \item[$(ii)$] the volume of the actual sessile droplet is conserved (Eq.\ (\ref{volG}));

 \item[$(iii)$] the total force on the \textit{full} droplet vanishes in the case of a pinned contact line; in the case of a free contact line the force balance is automatically satisfied by fixing the center of mass. 
\end{itemize}
%\vspace*{0.5cm}

Condition $(iii)$ represents a sufficient condition for mechanical equilibrium of the sessile droplet, because it implies that every piece of the full spherical interface is in equilibrium. In fact, conditions $(ii)$ and $(iii)$ are automatically satisfied for the Green's functions which obey Eqs.\ (\ref{govGN}) and (\ref{govGD}), because those equations have been actually obtained under the conditions of force balance and volume constraint. However, the physical assumptions as formulated in $(ii)$ and $(iii)$ can provide a guide for finding the  distribution of images. The boundary condition expressed in $(i)$ has to be imposed separately.

According to the above reasoning, the Green's functions for $\theta_0=\pi/2$ can be written in the following general form:
\begin{equation}
	G_{\sigma}(\Omega,\Omega')= G(\Omega,\Omega')+\sum_{i}Q_{\sigma i}G(\Omega,\Omega_{\sigma i}) + G_{\sigma,corr}(\Omega,\Omega'),
	\label{ND_greens}
\end{equation}

\noindent where $G(\Omega,\Omega')=G(\Omega',\Omega)$ (see, c.f., Eq.\ (\ref{greens_function})) is the Green's function for a \textit{free} droplet (see, c.f., Eq.\ (\ref{greens})). Both the amplitudes $Q_{\sigma i}$ and the directions $\Omega_{\sigma i}$ of the images, as well as a possible correction $G_{\sigma,corr}$ must be chosen such that the conditions $(i)-(iii)$ and the Green's equations (\ref{govGN}) and (\ref{govGD}) are satisfied. The problem of a single pointlike force acting at the surface of a \textit{free} droplet with a fixed center of mass has been studied by Morse and Witten in Ref.~\cite{Morse1993}, who derived the following equation for the free $G$:
\begin{equation}
	-(\nabla_a^2+2)G(\Omega,\Omega')=\sum_{l\geq 2,m}Y^*_{lm}(\Omega)Y_{lm}(\Omega')=\hat{\delta}(\Omega,\Omega'),
	\label{greens}
\end{equation}

\noindent where the rhs is a modified delta distribution $\hat{\delta}(\Omega,\Omega')$ with the components $l=0$ and $l=1$ projected out, {so that $G(\Omega,\Omega')=\sum_{l\geq 2,m}(l(l+1)-2)^{-1}Y^*_{lm}(\Omega)Y_{lm}(\Omega')$. This reflects} the conditions of incompressibility of the liquid {($\int_{S_2}\!d\Omega\,G(\Omega,\Omega')=0$ where $S_2$ is the full unit sphere)} and of balance of forces acting on the droplet {($\int_{S_2}\!d\Omega\,\boldsymbol{e}_r\,G(\Omega,\Omega')=0$)}. The solution of Eq.\ (\ref{greens}) can be expressed in closed form as~\cite{Morse1993}
\begin{multline}
%\begin{equation}
G(\Omega,\Omega')\equiv G(\bar{\theta})=\\
=-\frac{1}{4\pi}\left[ \frac{1}{2} + \frac{4}{3}\cos\bar{\theta} + \cos\bar{\theta}\ln\left(\frac{1-\cos\bar{\theta}}{2}\right) \right],
\label{greens_function}
%\end{equation}
\end{multline}

\noindent where $\bar{\theta}$ is the angle between the unit vectors pointing into the directions $\Omega$ and $\Omega'$. Morse and Witten pointed out that the deformation $v$ of a full droplet can be described in terms of $G$ also in the case that the center of mass is not fixed but, instead, there are other pointlike forces distributed such that the total force on the droplet vanishes. We shall use this latter property of the free $G$ for determining the Green's function $G_B$ of a \textit{sessile} droplet with a pinned contact line.

As we shall show in the following subsections, in the case $\theta_0=\pi/2$ the method of images can be applied successfully because the reflection at the surface plane of the substrate provides a smooth interface between the actual and virtual droplets. In the case $\theta_0\neq\pi/2$ the interface has a non-physical cusp and the method cannot be applied directly. However, we cannot rule out the possibility that there is a modification of the method which can be used successfully also for $\theta_0\neq\pi/2$ (for the case $\alpha=0$ see Appendix C).

%%%%%%%%%%%%%%%%%%%%%%%%%%%%%%%%%%%%%%%%%%%%%%%%%%%%%%%%%%%%%%%%%%%%%%%%%%%%%%%%%%%%
\subsection{Free contact line}
In the case of a free contact line a sessile droplet would move along the substrate if exposed to a non-vanishing lateral component of the external force applied to the particle. Therefore, in order to achieve a motion-free equilibrium, for example the lateral position of the center of mass has to be fixed. Physically, this compensation can be achieved by applying a body force to the droplet. Fixing the center of mass is to a certain extent artificial, but provides the simplest model for a free contact line. (For example, if a particle would hit the droplet and the relaxation of the position of the center of mass would last much longer then the relaxation of capillary waves generated by the impact, then the inertia of the droplet would effectively fix the center of mass. Alternatively, in the presence of gravity and depending on $\alpha$ a small tilt of the substrate could provide a lateral body force counterbalancing the lateral component of $f\boldsymbol{e}_r$.)

\begin{figure}[ht]
	\centering
	\psfragscanon
	\psfrag{f}[l][l][1]{{\large$f$}}
	\psfrag{f1}[c][c][1]{{\large$-f\sin\alpha$}}
	\psfrag{alpha}[l][l][1]{{\large$\alpha$}}
	\psfrag{x}[l][l][1]{{\large$x$}}
	\psfrag{z}[l][l][1]{{\large$z$}}
	\includegraphics[width=0.42\textwidth]{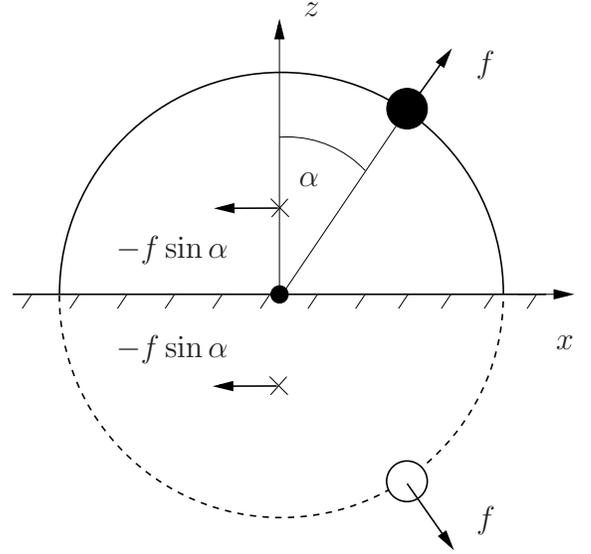}
	\psfragscanoff
	\caption{In the case of a free contact line the total drop, being the union of the actual (upper full line) and virtual (lower dashed line) drop, is pulled by $f$ in the upper part and by its mirror image in the lower part, which corresponds to the Green's function in Eq.\ (\ref{greens_N}) (with $\Omega'=\Omega_1=(\theta'=\alpha,\phi'=0)$). The lateral force of magnitude $-f\sin\alpha$ fixes the center of mass ($\times$) of each hemisphere separately.}
	\label{imageN}
\end{figure}

For $\theta_0=\pi/2$ the boundary condition in Eq.\ (\ref{boundary_N}) takes the simple form
\begin{equation}
  \partial_{\theta}G_A(\Omega,\Omega')\vert_{\Omega\in\partial\Omega_0}= 0,
\label{boundary_Npi2}
\end{equation}

\noindent which can be identified with the Neumann boundary condition. Equation (\ref{boundary_Npi2}) and the volume constraint $\int_0^{2\pi}\!d\phi\,\int_0^{\pi/2}\!d\theta\sin\theta\, G_A(\Omega,\Omega')=0$ can be satisfied by the Green's function given in Eq.\ (\ref{ND_greens}) with $Q_{A1}=1$ and $\Omega_{A1}=\hat{Z}\Omega'$ where $\hat{Z}$ means the reflection with respect to the plane $z=0$ (see Fig.~\ref{imageN}):
\begin{equation}
	G_A(\Omega,\Omega') = G(\Omega,\Omega')+G(\Omega,\hat{Z}\Omega').
	\label{greens_N}
\end{equation}

\noindent If the {\textit{full}} drop (union of the actual and of the virtual drop) is exposed to two point forces $\pi(\Omega)=\delta(\Omega,\Omega')+\delta(\Omega,\hat{Z}\Omega')$ its resulting shape, given by Eqs.\ (\ref{ND-solution}) and (\ref{greens_N}), renders the shape of the \textit{actual} droplet for $z>0$. For this ansatz, with $G(\Omega,\Omega')$ given by Eq.\ (\ref{greens_function}), it can be checked that $\partial_{\theta} G_A$ vanishes at the contact line. The correction term $G_{A,corr}$ vanishes, because the contributions to the volume stemming from the two terms in Eq.\ (\ref{greens_N}) cancel each other, so that the volume constraint is fulfilled for $G_{A,corr}\equiv0$. 

In order to discuss the force balance we consider, without loosing generality, $\phi'=0$. In such a case, due to the mirror symmetry with respect to the $xz$-plane and the mirror symmetry with respect to the $xy$-plane, the $y$-component and the $z$-component of the external force on the \textit{full} drop due to the two point forces vanish, whereas the $x$-component has the strength $2\sin\theta'$ (see Fig.\ \ref{imageN} with $\alpha=\theta'$ and $f=1$). Mechanical equilibrium in this direction must be restored by fixing the center of mass which corresponds to applying the forces of strength $Q_{CM}=-\sin\theta'$ (Eq.\ (\ref{qcm})) both to the \textit{actual} and to the \textit{virtual} drop.

Finally, by using Eq.\ (\ref{greens}) one can check explicitly that $G_A$ as given by Eq.\ (\ref{greens_N}) satisfies Green's equation for $\theta_0=\pi/2$.

%%%%%%%%%%%%%%%%%%%%%%%%%%%%%%%%%%%%%%%%%%%%%%%%%%%%%%%%%%%%%%%%%%%%%%%%%%%%%%%%%%%%
\subsection{Pinned contact line}
In this subsection we consider the case that the contact line is pinned at the circle corresponding to the reference configuration. One can propose that this pinning can be accomplished by substrate heterogeneities; but this requires a dedicated design in order to maintain the circular shape of the contact line. This model has the virtue that in contrast to the previous one the balance of forces is automatically satisfied without fixing the center of mass.

\begin{figure}[ht]
	\centering
	\psfragscanon
	\psfrag{f}[l][l][1]{{\large$f$}}
	\psfrag{f1}[c][c][1]{{\large$-f$}}
	\psfrag{f2}[l][l][1]{{\large$2f\cos\alpha$}}
	\psfrag{alpha}[l][l][1]{{\large$\alpha$}}
	\psfrag{x}[l][l][1]{{\large$x$}}
	\psfrag{z}[l][l][1]{{\large$z$}}
	\includegraphics[width=0.45\textwidth]{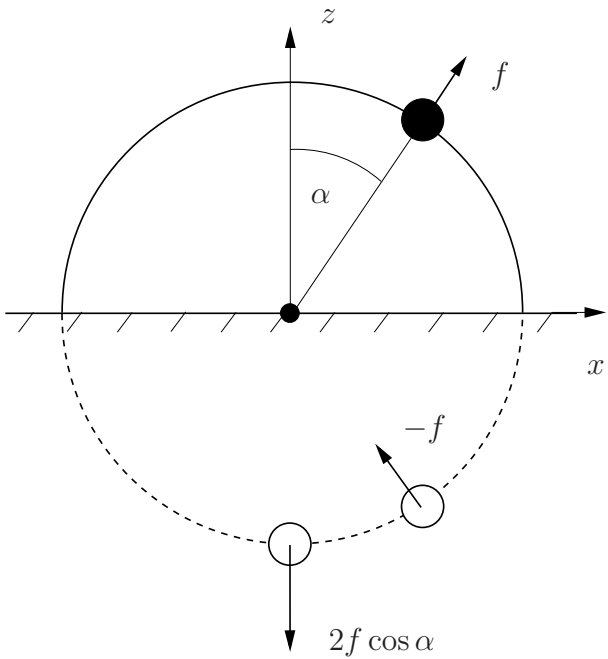}
	\psfragscanoff
	\caption{Distribution of images in the case of a pinned contact line, corresponding to the Green's function $G_B$ in Eq.\ (\ref{greens_D}) (with $\Omega'=\Omega_1=(\theta'=\alpha,\phi'=0)$). The correction $G_{B,corr}$ (Eq.\ (\ref{g_corr})) does not give rise to any forces. The center of mass is free.}
	\label{imageD}
\end{figure}

By choosing $Q_{B1}=-1$ positioned at $\Omega_{B1}=\hat{Z}\Omega'$ the boundary condition in Eq.\ (\ref{boundary_D}) is automatically satisfied. However, the corresponding pressure field for pointlike forces placed at $\Omega'$ and at $\hat{Z}\Omega'$, which corresponds to $\pi(\Omega)=\delta(\Omega,\Omega')-\delta(\Omega,\hat{Z}\Omega')$, leads to a non-vanishing net force acting on the droplet as a whole, which is directed vertically and is equal to $2\cos\theta'$ (see Fig.\ \ref{imageD}). In order to restore the force balance (without fixing the center of mass) we place a second image $Q_{B2}=2\cos\theta'$ at the bottom ("south pole", $\theta=\pi$) of the virtual droplet, so that $\Omega_{B2}=\Omega_{\pi}$ (see Fig.\ \ref{imageD} with $f=1$ and $\alpha=\theta'$). However, this second image contributes to a  deformation at $\theta=\theta_0=\pi/2$, which violates the boundary condition in Eq.\ (\ref{boundary_D}). This additional contribution, which is equal to $2(\cos\theta')G(\Omega=(\theta=\pi/2,\phi),\Omega_{\pi})=: -I(\theta')$ and thus depends neither on $\Omega$ nor on $\phi'$, has to be subtracted from the Green's function in order to uphold Eq.\ (\ref{boundary_D}). Finally, the volume constraint is fulfilled by adding a second term $H(\Omega')\cos\theta$, which corresponds to a rigid vertical translation {(i.e., in $z$ direction)} of the droplet and to an effective pressure $-(\nabla_a^2+2)H(\Omega')\cos\theta=-H(\Omega')(\nabla_a^2+2)\cos\theta$ which vanishes independently of $H(\Omega')$ {(which is in accordance with a rigid translation)}. Thus, this term does not contribute to the net force on the droplet. Moreover, it vanishes for $\theta=\pi/2$, i.e., at the contact line, so that it does not violate the boundary condition in Eq.\ (\ref{boundary_D}). Accordingly, we are led to the following ansatz for the correction term: 
\begin{equation}
	G_{B,corr}(\Omega,\Omega')=H(\Omega')\cos\theta+I(\theta'),
\label{g_corr}
\end{equation}

\noindent so that the ansatz for the total Green's function is
\begin{multline}
%\begin{equation}
	G_B(\Omega,\Omega') = G(\Omega,\Omega')-G(\Omega,\hat{Z}\Omega')+2G(\Omega,\Omega_{\pi})\cos\theta'\\ +H(\Omega')\cos\theta+I(\theta').
\label{greens_D}
%\end{equation}
\end{multline}

\noindent Inserting the above expression for $G_B$ into the boundary condition in Eq.\ (\ref{boundary_D}) renders 
\begin{equation}
	I(x)=\dfrac{\cos x}{4\pi}.
	\label{Ax}
\end{equation} 

\noindent The volume constraint in Eq.\ (\ref{volG}) provides an expression for $H(\Omega')= H(\theta')$:
\begin{equation}
	H(x)=\frac{1}{2\pi}\left[\cos x\ln\left(\frac{1+\cos x}{2}\right)-\cos x\right].
\label{Bx}
\end{equation}

\noindent Finally, by using Eq.\ (\ref{Ax}), it can be checked explicitly that $G_B$ as given by Eq.\ (\ref{greens_D}) satisfies the Green's equation (\ref{govGD}) for $\theta_0=\pi/2$; the contribution to Eq.\ (\ref{govGD}) due to $H(\Omega')$ (see Eq.\ (\ref{greens_D})) vanishes because $(\nabla_a^2+2)\cos\theta=0$. 

%%%%%%%%%%%%%%%%%%%%%%%%%%%%%%%%%%%%%%%%%%%%%%%%%%%%%%%%%%%%%%%%%%%%%%%%%%%%%%%%%%%%
\subsection{Free energy}
As discussed in Subsec.\ III.D the excess free energy $\Delta F_{\sigma}$ is well-defined even in the limiting case that $\pi(\Omega)$ is of the form $\pi(\Omega)=q\,\delta(\Omega,\Omega_1)$. In this case and for $\theta_0=\pi/2$ one obtains the closed expression (Eqs.\ (\ref{DFs}), (\ref{greens_N}) and (\ref{greens_D}))
\begin{align}
	\Delta F_{\sigma}(\alpha) &= -\frac{f^2}{2\gamma}[g_{\sigma}(\alpha)-g_{\sigma}(0)], \label{excess_free_enND}
\end{align}

\noindent where the functions $g_A$ and $g_B$ are both regular at $\alpha=0$ and given explicitely by
\begin{equation}	
	g_A(\alpha)= G(\bar{\theta}=\pi-2\alpha) \label{gN}
\end{equation}

\noindent and
\begin{multline}
	g_B(\alpha)= -G(\bar{\theta}=\pi-2\alpha) + 2G(\bar{\theta}=\pi-\alpha)\cos\alpha\\
	 + H(\alpha)\cos\alpha + I(\alpha), \label{gD}
\end{multline}

\noindent with $G(\bar{\theta})$ given by Eq.\ (\ref{greens}); the functions $I(\alpha)$ and $H(\alpha)$ are given by Eqs.\ (\ref{Ax}) and (\ref{Bx}), respectively.

%%%%%%%%%%%%%%%%%%%%%%%%%%%%%%%%%%%%%%%%%%%%%%%%%%%%%%%%%%%%%%%%%%%%%%%%%%%%%%%%%%%%%
%%%%%%%%%%%%%%%%%%%%%%%%%%%%%%%%%%%%%%%%%%%%%%%%%%%%%%%%%%%%%%%%%%%%%%%%%%%%%%%%%%%%%

\section{Numerical results}

In this section we compare the analytical expressions for $\Delta F_A$ and $\Delta F_B$, given within perturbation theory by Eqs.\ (\ref{excess_free_enND})-(\ref{gD}) and (\ref{greens_function}) for point force induced deformations, with the numerical minimization of $\Delta F$ defined by Eqs.\ (\ref{functional})-(\ref{excess_free_en}) for a deformation induced by a particle of finite extent $a$ (Fig.~\ref{sketch}) and for $\theta_0=\pi/2$, employing a finite element method~\cite{Brakke}. The contact line at the particle is taken to be free which leads to a fixed contact angle $\theta_p$ at the particle, in analogy to a free contact line at the substrate which fixes the contact angle $\theta_0$ at the substrate.
In the case of axially symmetric configurations we compare the numerical results with the exact expressions given in Appendix B.

%%%%%%%%%%%%%%%%%%%%%%%%%%%%%%%%%%%%%%%%%%%%%%%%%%%%%%%%%%%%%%%%%%%%%%%%%%%%%%%%%%%%
\subsection{Axially symmetric configurations with free contact line}
First, we study the axially symmetric configuration with the particle positioned at the apex of the droplet. For this particular configuration the free energy can be calculated analytically solving the full non-linear Young-Laplace equation, i.e., without linearizing in terms of $\epsilon$ (Appendix B). In this case it is more suitable to choose the displacement $h$ instead of the force $f$ as the independent variable. The free energy $F_{A0}(f)$ for $\alpha=0$, as introduced in Eq.\ (\ref{excess_free_en}) and in the text preceding Eq.\ (\ref{excess_free_en}), can be expressed as the Legendre transform of the free energy $\tilde{F}_{A0}(h)$ given implicitly by
\begin{equation}
	F_{A 0}(f,\theta_0,\theta_p,a,R_0)=\min_{h}[\tilde{F}_{A0}(h,\theta_0,\theta_p,a,R_0)-fh].
	\label{free_en_lagrange}
\end{equation}

\begin{figure}
	%\begin{flushleft}	%
	\centering
	\psfragscanon
	\psfrag{h}[c][c][1.2]{$h/a$}
	\psfrag{Fh}[c][c][1.2]{$\tilde{F}_{A0}(h)/(\gamma a^2)$}
	\psfrag{a}[c][c][1.2]{$\alpha=0$}
	\psfrag{x1}[c][c][1.2]{$h/a$}
	\psfrag{y1}[c][c][1.2]{$\tilde{f}(h)/(\gamma a)$}
	%\hspace*{-6.2cm}
	\begin{overpic}[width=0.45\textwidth]{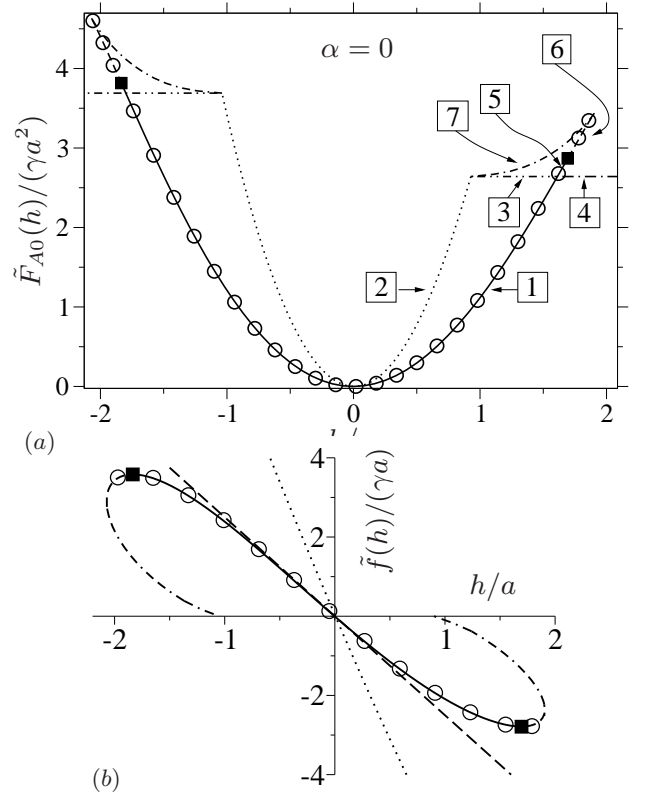}
	\put(0,0){$(a)$}
	%\put(105,13){$(b)$}
	\end{overpic}\\
	\begin{overpic}[width=0.35\textwidth]{Fig07}
	\put(0,0){$(b)$}
	\end{overpic}
	\psfragscanoff
	\caption{ The surface free energy $\tilde{F}_{A0}(h)$ $(a)$ for axially symmetric configurations ($\alpha=0$) with $\theta_0=\pi/3, \theta_p=\pi/2$, and $V_l=79\frac{4\pi}{3}a^3$ and the total capillary force $\tilde{f}(h)=-\partial\tilde{F}_{A0}/\partial h$ $(b)$ as functions of immersion $h$. For $h>0$ and $h<0$ one can distinguish 7 branches (for reasons of clarity only the branches for $h>0$ have been tagged in the plot). Branch 1 corresponds to the globally stable, full analytic solution of the Young-Laplace equation. Branch 2 is the free energy under the constraint that the droplet keeps the shape of a spherical cap, which is the equilibrium configuration for $h=0$. For $h\neq0$ this constraint induces $\theta_p$ to deviate from its value $\pi/2$. Lifting this constraint lowers the free energy towards the globally stable branch 1, on which $\theta_p$ has its fixed value $\pi/2$. For $h/a>0.93$ and $h/a<-1.04$ the particle detaches from the interface if one keeps the droplet shape to be a spherical cap (3, 4). On branch 3 (4) this detached configuration is metastable (globally stable). The extension of branch 1 beyond the horizontal branches 3 and 4 is metastable on branch 5 up to the filled square, which indicates an inflection point. Beyond that point branch 6 is unstable. Branch 7 is an extension of branch 2 which is metastable with respect to detachment, i.\ e., branches 3 and 4. Branches 6 and 7 merge at the point at which the unstable solution along branch 6 ceases to exist. The equilibrium configurations are given by the branches 1 and 4. Lines correspond to analytic results whereas circles correspond to numerical finite element calculations. The various line codes in $(a)$ and $(b)$ correspond to each other. In $(b)$ the straight long-dashed line indicates the slope of the analytic solution at $h=0$. The force maximum $\tilde{f}_{max}/(\gamma a)=3.57$ occurs at $h/a=-1.82$ and the minimum $\tilde{f}_{min}/(\gamma a)=-2.79$ at $h/a=1.70$.}
	\label{symmetric}
	%\end{flushleft}
\end{figure}

\noindent In the following, when referring to ``stability'' of the branches of the free energy (Fig.\ \ref{symmetric}) we shall actually mean the stability of the configurations with an external force $f$ acting on the particle (with $f=\partial\tilde{F}_{A0}/\partial h$, see also Appendix C). Another choice for the independent variable might be the angular position of the contact line at the particle parameterized by the polar angle $\beta$ (see, c.f., Fig.~\ref{axisymmetric}$(b)$ in Appendix B). In this case and for a free contact line at the substrate the expression for the free energy in terms of elliptic functions is derived in Appendix B. We have carried out finite element calculations also for $\tilde{F}_{A0}(h)$ so that the comparison with the analytic results serves as a welcome test for the performance of the numerical code we have to rely on for configurations which are not axially symmetric. (Note that for $\alpha=0$ there is no need to fix the lateral position of the center of mass.) In this case the numerical procedure consists of employing the finite element method~\cite{Brakke} for the minimization (for details see Subsec.V.B) of the free energy functional in Eq.\ (\ref{functional}) with $\alpha=0$, $f=0$, and for fixed $h$. We have found very good agreement between the numerical and analytical results (see Fig.~\ref{symmetric}).

In Fig.~\ref{symmetric} we present the results for $\theta_0=\pi/3, \theta_p=\pi/2$, and $R_0/a\approx 8$. One can distinguish seven branches of the free energy both for $h<0$ and for $h>0$. The metastable branches denoted as $7$ lead to a maximal value of $\tilde{F}_{A0}$ on each side of the curve and they correspond to configurations with the contact line close to one of the poles of the particle. Once the system is prepared in such a configuration at the upper part of branch 7 and the constraint of fixed $h$ is lifted, the particle slides down branch 7 by shrinking the contact line while moving closer to the reference interface $h=0$, ending up in a state in which it is fully immersed in the liquid or in the gas phase. 

The filled squares in Fig.~\ref{symmetric} indicate inflection points of the free energy curve. According to the following reasoning, for a large droplet these points are located (for $\theta_p=\pi/2$) close to $\beta =\pi/2-\pi/4= \pi/4$, which corrresponds to $h/a=-1.91$, and $\beta=\pi/2+\pi/4=3\pi/4$, which corrresponds to $h/a=1.60$. By definition, at an inflection point $\partial \tilde{F}_{A0}/\partial h$ is extremal, which implies that the capillary force exerted on the particle by the interface, defined as
\begin{equation}
 \tilde{f}(h)=-\dfrac{\partial\tilde{F}_{A0}}{\partial h}=-f(h), \label{tildef}
\end{equation}

\noindent also has an extremum. However, instead of discussing the derivative, which becomes complicated as soon as $\tilde{F}_{A0}(h)$ is given implicitly by a set of equations (see Appendix B), we resort to a physical picture. 

The total capillary force has two components. The first one is the force acting on the contact line at the particle, which for a unit length $dl$ of the contact line, equals $\gamma \boldsymbol{e}_t dl$, where $\boldsymbol{e}_t$ is a unit vector tangential to the liquid-gas interface and normal to the contact line, pointing in the direction away from the particle (see, e.g., Ref.~\cite{Segel_book}). In the case of an axisymmetric interface, integrating over the contact line leaves only the vertical component, equal to $-2\pi\gamma a \sin\beta\sin\psi_0$, where $2\pi a\sin\beta$ is the circumference of the three-phase contact line, the tangent of the angle $-\psi(r)$ equals the slope of the interface at a distance $r$ from the axis, and, according to  Young's law, $\psi_0=\psi(r=a\sin\beta)=\beta-\theta_p$ (see, c.f., Appendix B and Fig.~\ref{axisymmetric}). The second contribution to the total capillary force stems from the internal pressure, which for an almost spherical interface is approximately equal to the Laplace pressure $2\gamma/R_0$ (given exactly by Eq.\ (\ref{lagrange})), resulting in a force of magnitude approximately $\pi(a \sin\beta)^22\gamma/R_0$ directed upwards, which is of the order $O(a/R_0)$ relative to the force on the contact line and therefore it can be neglected for $R_0\gg a$. Hence, for $\theta_p=\pi/2$ (as studied in Fig.\ \ref{symmetric}) the force on the particle equals approximately $\pi\gamma a \sin(2\beta)$, with its extremes approximately being $\pi\gamma a$ and $-\pi\gamma a$ at {$\beta=\pi/4=\pi/2-\pi/4$ and $\beta=3\pi/4=\pi/2+\pi/4$,} respectively. 

\begin{figure}
\begin{flushleft}
 	\psfragscanon
	\psfrag{be1}[l][l][1]{\large{$\beta=\pi/4$}}
	\psfrag{be2}[l][l][1]{\large{$\beta=3\pi/4$}}
	\psfrag{r}[c][c][1]{\large{$r/a$}}
	\psfrag{z}[c][c][1]{\large{$\dfrac{z}{a}$}}
	\psfrag{th}[c][c][1]{\large{$\theta$}}
	\psfrag{u}[c][c][1]{\large{$u(\theta)$}}
	\psfrag{a}[c][c][1]{\large{$a$}}
	\psfrag{hl}[r][r][1]{\large{$-1.91=\dfrac{h}{a}$}}
	\psfrag{hg}[r][r][1]{\large{$1.6=\dfrac{h}{a}$}}
	\hspace*{1cm}
	\begin{overpic}[width=0.35\textwidth]{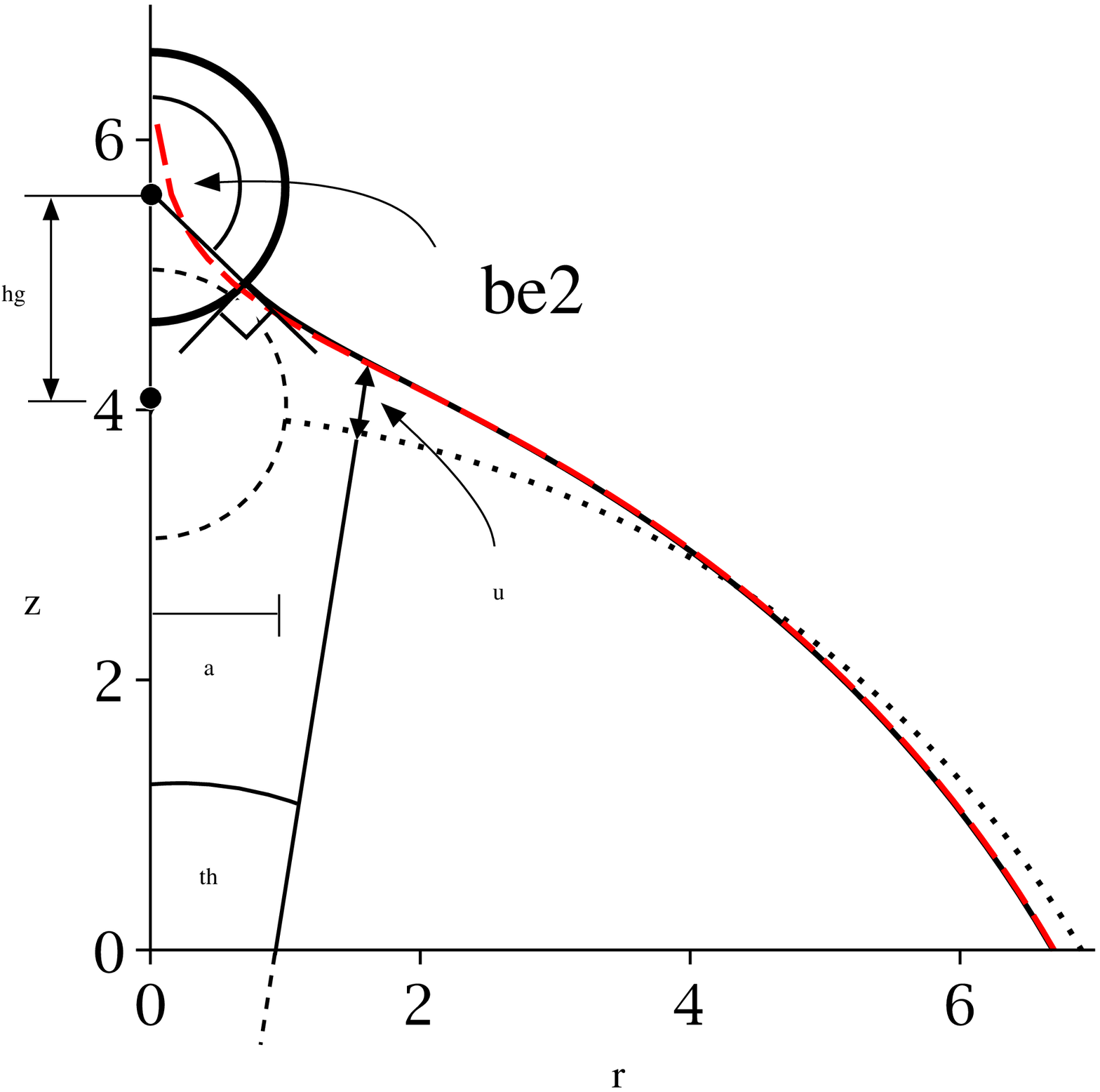}%{profile_th60_a17}
	\put(75,90){{\large$(a)$}}		 
	\end{overpic}
	\vspace*{0.7cm}
	\hspace*{0.7cm}
	\begin{overpic}[width=0.35\textwidth]{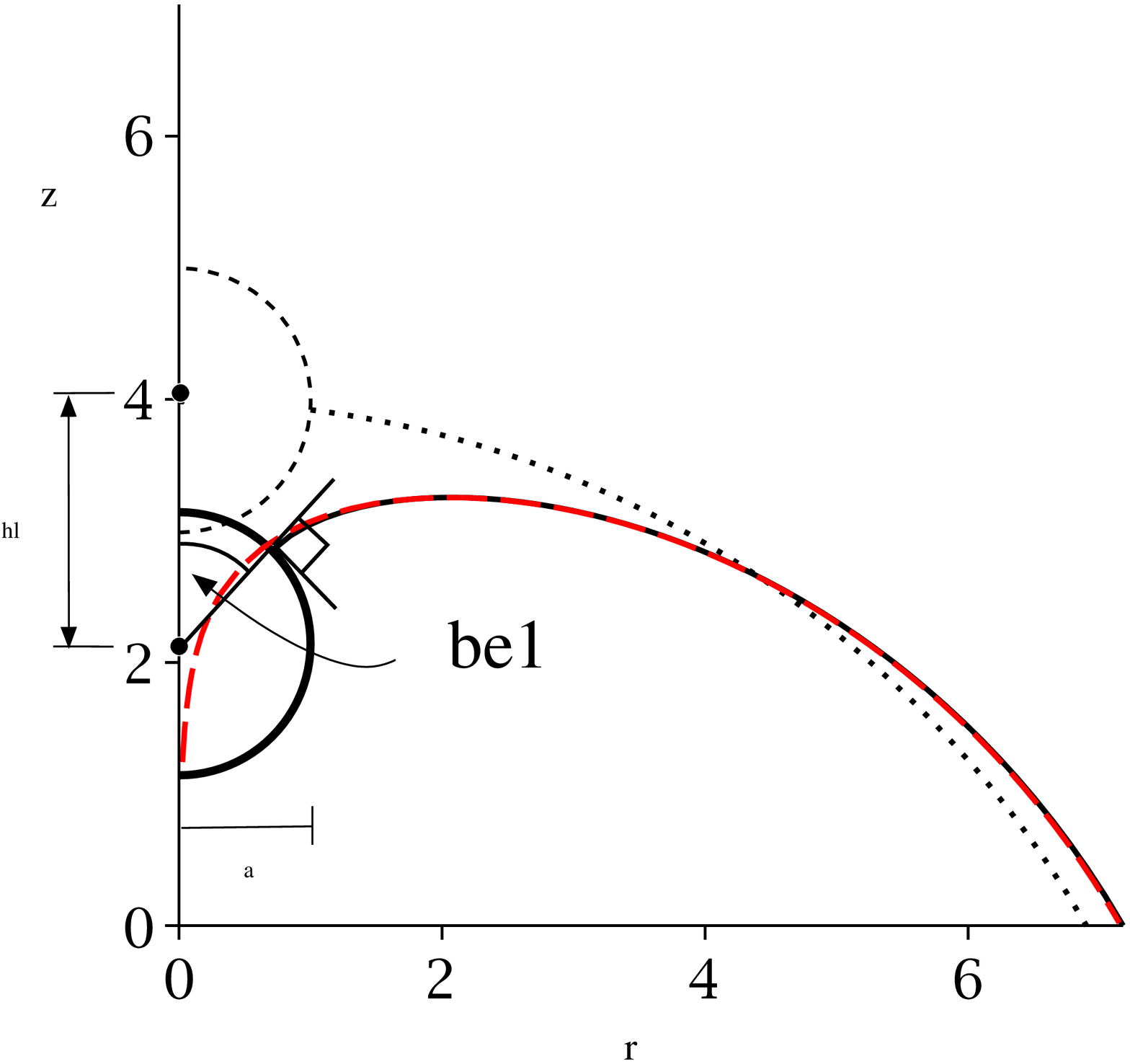}
	\put(75,90){\large{$(b)$}}	
	\end{overpic}
	\psfragscanoff
	\caption{Droplet shapes for $\theta_p=\pi/2$ and fixed $\theta_0=\pi/3$ (model $A$), calculated by using the exact solution of the capillary equation (\ref{young-laplace}) (black solid lines beneath the red dashed lines) for $\beta=3\pi/4$ in $(a)$ and for $\beta=\pi/4$ in $(b)$, and by using the approximate solutions $R_0\epsilon v_0(\theta)/a=u(\theta)/a$ (red dashed lines) given by Eqs.\ (\ref{v_symmetric})-(\ref{Q0}) for pointlike forces of amplitude $f/(\gamma a)=2.77$ and $f/(\gamma a)=-3.56$, respectively (compare Fig.\ \ref{symmetric}$(b)$ and Appendix C), with a cut-off at $\theta=0.005$ and with the remaining set of parameters the same as in Fig.~\ref{symmetric}. We note the logarithmic divergence of the approximate solutions at $\theta=0$. The dotted lines correspond to the particle and the spherical cap shapes of the droplet in the reference configuration.}
	\label{symmetric_analytic}
\end{flushleft}
\end{figure}

As a consequence of the liquid volume constraint, even for large droplets the free energy $\tilde{F}_{A0}(h)$ must be always slightly asymmetric with respect to $h=0$ {(or equivalently $\beta=\pi/2+O(a/R_0)$)}. 
This can be understood by comparing the two most extreme configurations. In order to maintain a constant volume the radius of the droplet for the configuration with the particle completely immersed in the liquid phase ($h/a=-1.06$ on branch 7, see Fig.\ \ref{symmetric}$(a)$) must be larger than for the configuration with the particle completely immersed in the gas phase ($h/a=0.94$). Because in these two cases the droplet shapes are spherical caps, the free energy is higher for the configuration with a larger droplet radius, i.e., for the one with the particle completely immersed in the liquid (as long as $\theta_p=\pi/2$ the surface energy at the particle does not vary with $h$). Accordingly, the total capillary force acting on the particle $\tilde{f}(h)$ is not exactly antisymmetric with respect to $h=0$. In the studied case of a medium-sized droplet the extremal values of the force, calculated by a numerical differentiation of $\tilde{F}_{A0}(h)$, equal $\tilde{f}_{max}/(\gamma a)=3.57$ and $\tilde{f}_{min}/(\gamma a)=-2.79$ for $h/a=-1.82$ and $h/a=1.70$, respectively (see Fig.\ \ref{symmetric}$(b)$); thus this  effect is clearly detectable. From the deviation of the full line from the dashed one in Fig.\ \ref{symmetric}$(b)$ one can also infer that the nonlinearity of $\tilde{f}(h)$ is stronger for $h>0$, i.e., for negative $\tilde{f}$. The deviation from the linear behavior sets in approximately for $\tilde{f}/(\gamma a)>2$ and for $\tilde{f}/(\gamma a)<-1$. 

In Fig.~\ref{symmetric_analytic} we compare the shapes of the droplet for the configurations with $\beta=\pi/4$ (for which $\tilde{f}/(\gamma a)=3.56$) and $\beta=3\pi/4$ (for which $\tilde{f}/(\gamma a)=-2.77$) calculated by using the exact solution of the non-linear Young-Laplace equation (Appendix B) and from linear perturbation theory. (For the case of axial symmetry Appendix C provides the solution of the linearized Young-Laplace equation also for $\theta_0\neq \pi/2$). Apparently, even close to  the edge of the stability regime the discrepancies between the two approaches are so small (actually $\beta=\pi/4$ corresponds already to a metastable configuration, so that it belongs to branch 6 for $h<0$ in Fig.\ \ref{symmetric}$(a)$), that they are practically beyond the resolution of Fig.\ \ref{symmetric_analytic};  we expect them to remain small also for the configurations without axial symmetry.

\subsection{Free energy of configurations without axial symmetry}
In order to determine numerically the dependence of the free energy on $\alpha$ we fix the angular position $\alpha$ of the particle. For such a configuration we minimize the free energy functional in Eq.\ (\ref{functional}) with respect to the radial displacement $h$ of the particle and thus the shape of the droplet by using a finite element method~\cite{Brakke}. This method tracks the evolution of an arbitrarily shaped initial body of liquid towards its shape corresponding to the minimum of the free energy. In order to obtain suitable start configurations with $\alpha\neq 0$ the surface is pre-evolved for $\alpha=0$ from a cubic shape and then the particle is moved away from the apex in a step-wise fashion by small increments $\delta\alpha$. In both model $A$ and model $B$ the contact line at the particle is taken to be free. In the case of model $B$ the contact line at the substrate is pinned from the very beginning at a circle corresponding to the reference configuration with $f=0$; accordingly, the geometrical center of the reference droplet (see Fig.\ \ref{sketch}), from which the radial distance of the particle is measured, is well defined throughout the evolution. In the case of model $A$ the situation is more complicated because, due to the finite size of the particle, the position $x_{CM,ref}$ of the center of mass of liquid depends on $\alpha$. This finite-size effect has to be taken into account for medium-sized droplets (see, c.f., the end of Subsec.\ V.B.2).

\begin{figure}[ht]
	%\vspace*{1cm}
	\centering
	\psfragscanon
	\psfrag{x1}[c][c][1]{}
	\psfrag{a1}[l][l][1]{$\alpha=72$ [degrees]}
	\psfrag{a2}[l][l][1]{$\alpha=48$ }
	\psfrag{a3}[l][l][1]{$\alpha=24$ }
	\psfrag{a4}[l][l][1]{$\alpha=0$ }
	\psfrag{r4}[l][l][1]{$R_0/a=4$ }
	\psfrag{r8}[l][l][1]{$\bigg\}\, 8$ }
	\psfrag{r12}[l][l][1]{$12$}
	\begin{overpic}[width=0.35\textwidth]{Fig10}
	\put(15,15){$(a)$}
	\put(-5,25){\rotatebox{90}{$\gamma\Delta F_B/f^2$}}
	\put(45,-3){$\alpha$ [degrees]}
	\end{overpic}\\
	\vspace*{0.5cm}
	\begin{overpic}[width=0.35\textwidth]{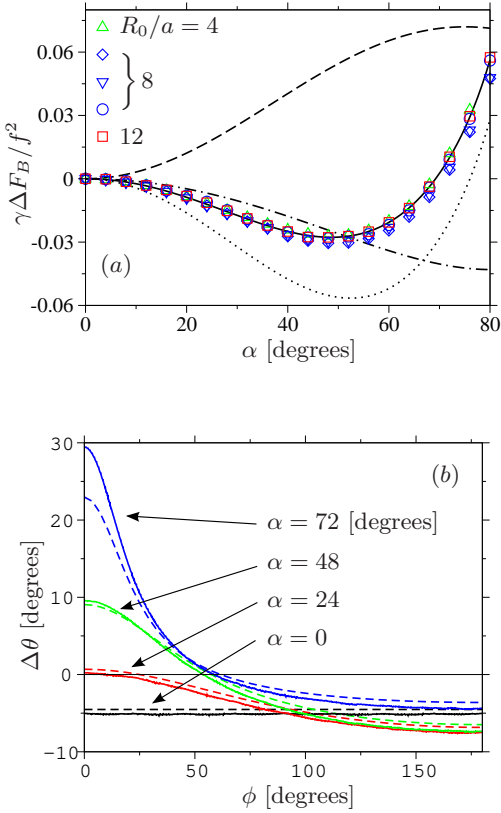}
	\put(85,65){$(b)$}
	\put(-2,25){\rotatebox{90}{$\Delta\theta$ [degrees]}}
	\put(45,-3){$\phi$ [degrees]}
	\end{overpic}
	\psfragscanoff
	\caption{$(a)$ Rescaled excess free energy (Eq.~(\ref{excess_free_en})) for a pinned contact line on the substrate as a function of the polar angle $\alpha$ (Fig.~\ref{sketch}) for a contact angle $\theta_0=\pi/2$ of the reference configuration at the substrate. The solid black curve corresponds to the analytic expression given by Eqs.\ (\ref{Ax})-(\ref{gD}). The dotted line and the dashed line are the contributions from the first and the second image (Fig.~\ref{imageD}), respectively, and the dash-dotted line is the volume correction (Eq.~(\ref{g_corr})). Green triangles ($\bigtriangleup$), blue circles and red squares correspond to $\{R_a/a=4,f/(\gamma a)=2,\theta_p=\pi/2\}$, $\{R_0/a=8,f/(\gamma a)=1,\theta_p=2\pi/3\}$, and $\{R_0/a=12,f/(\gamma a)=2,\theta_p=\pi/2\}$, respectively. Blue diamonds and inverted triangles ($\bigtriangledown$) corespond to $\{R_0/a=8,f/(\gamma a)=-1.5,\theta_p=\pi/2\}$ and $\{R_0/a=8,f/(\gamma a)=-2,\theta_p=\pi/2\}$, respectively. $(b)$ The difference $\Delta\theta(\phi)=\tilde{\theta}(\phi)-\theta_0$ between the actual contact angle $\tilde{\theta}(\phi)$ and $\theta_0=\pi/2$ as a function of the azimuthal angle $\phi\in[0,\pi]$ for $\{R_a/a=4,f/(\gamma a)=2,\theta_p=\pi/2\}$ and for various angular positions $\alpha$ of the particle. The dashed lines correspond to the approximate analytic expression (Eq.\ (\ref{tildetheta})) and the solid lines are numerical results. The thin horizontal line at $\Delta\theta=0$ corresponds to the reference configuration.}
	\label{dirichlet}
\end{figure}

\subsubsection{Pinned contact line on the substrate}
In Fig.~\ref{dirichlet}$(a)$ we plot numerical values of $\gamma\Delta F_B/f^2$ for various droplet sizes $R_0$, both signs and various strengths of $f$, and various contact angles $\theta_p$ at the particle. The variation of all these parameters does not affect the $\alpha$-dependence and we obtain a single master curve, in very good agreement with the theoretical expression in Eq.\ (\ref{excess_free_enND}) for a pointlike force. For the latter the contributions from both images (dashed line and dotted line) and from the volume correction $G_{B,corr}$ (dash-dotted line) are all equally important. We emphasize the de facto independence of $\Delta F$ from the contact angle $\theta_p$ at the particle (see Fig.\ 7 and additional data not shown), which justifies our perturbation theory which does not take into account the specific wetting energy of the particle. The only systematic deviation from the analytic expression occurs for negative value of $f$. We have investigated the system for various absolute values of negative $f$ (data not shown) and the free energy turns out to be always slightly overestimated by the analytic theory. In fact we obtain two master curves (for fixed $R_0/a$), which are close to each other: one for positive and one for negative $f$. 
\begin{figure}[ht]
	\centering
	\psfragscanon
	\psfrag{t0}[c][c][1]{$\theta_0$[degrees]}
	\psfrag{fmin}[c][c][1]{$\gamma\Delta F_{B,min}/f^2$}
	\psfrag{amin}[c][c][1]{$\alpha_{min}$[degrees]}
	\psfrag{pi6}[c][c][1]{$\pi/6$}
	\psfrag{pi3}[c][c][1]{$\pi/3$}
	\psfrag{pi2}[c][c][1]{$\pi/2$}
	\psfrag{2pi3}[c][c][1]{$2\pi/3$}
	\psfrag{5pi6}[c][c][1]{$\theta_0=5\pi/6$}
	\psfrag{f}[c][c][1]{$\gamma\Delta F_B/f^2$}
	\psfrag{a}[c][c][1]{$\alpha$[degrees]}
	\hspace*{-0.5cm}
	\begin{overpic}[width=0.35\textwidth]{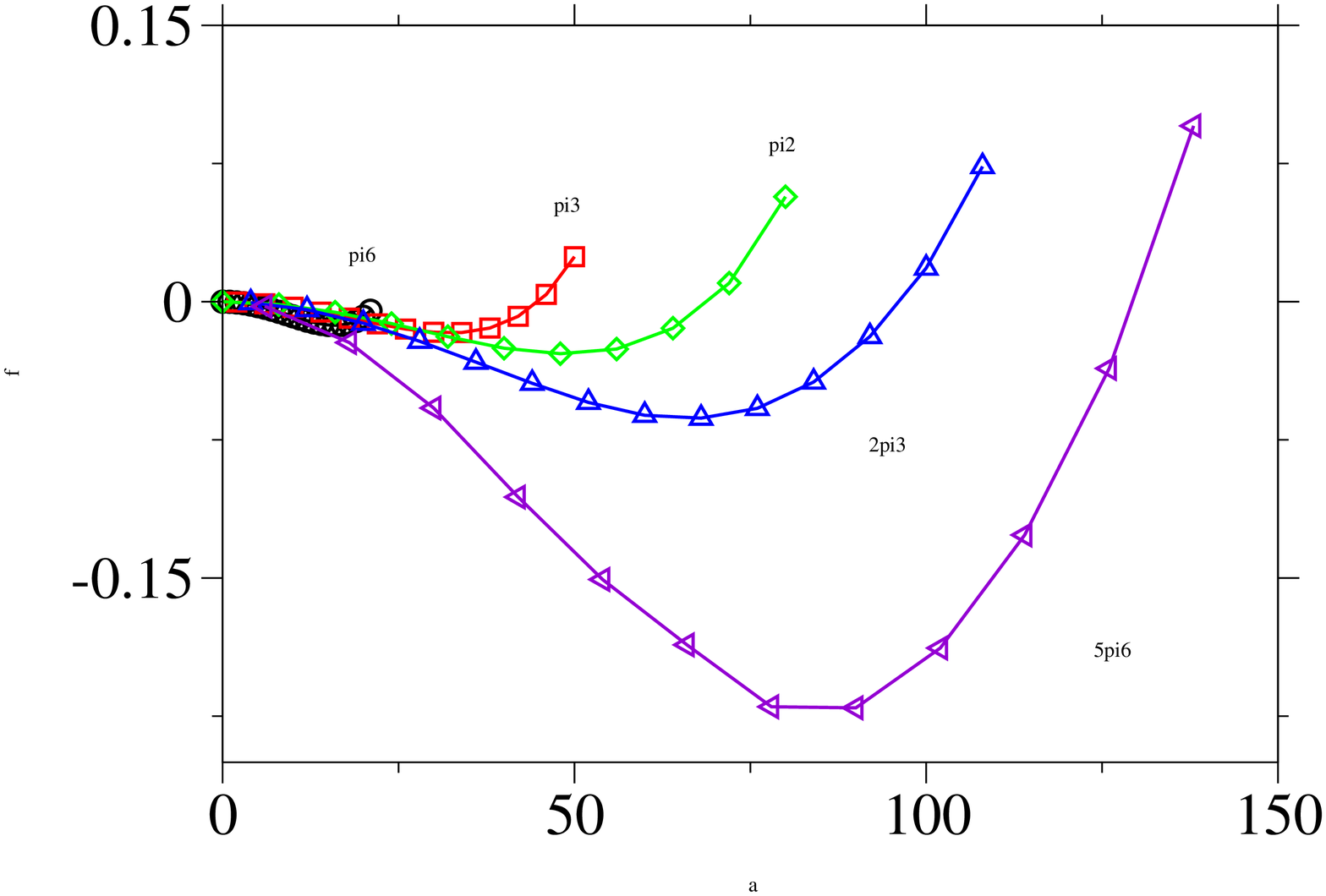}%{c1_fix_theta7}
	\put(18,13){$(a)$}
	\end{overpic}\\
	\vspace*{0.5cm}
	\begin{overpic}[width=0.37\textwidth]{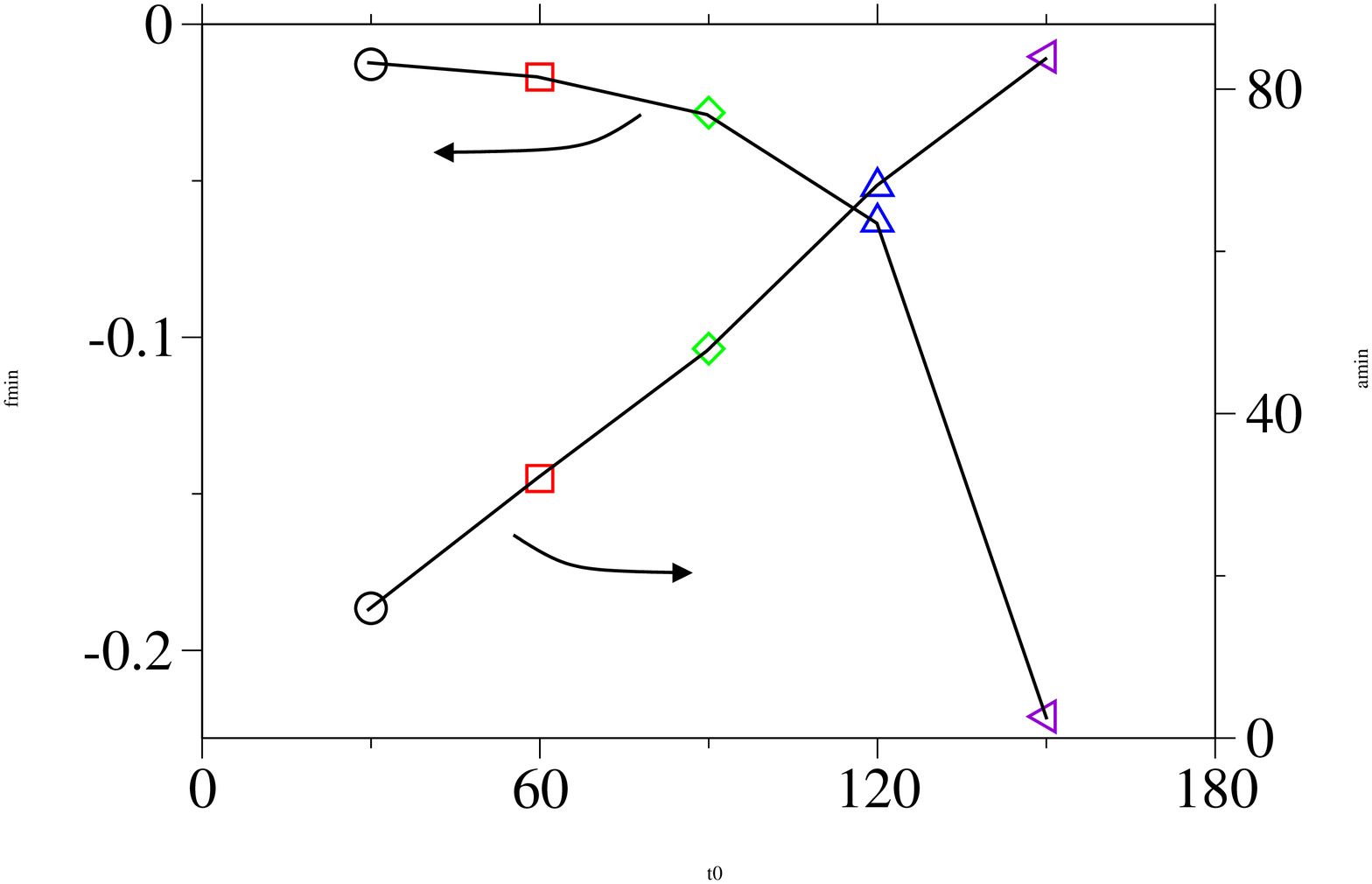}%{c1_fix_theta_min7}
	\put(16,13){$(b)$}
	\end{overpic}
	\psfragscanoff
	\caption{$(a)$ Numerical results for the dependence of the rescaled excess free energy (Eq.\ (\ref{excess_free_en})) on the angular position of the particle for various values of the contact angle $\theta_0$ of the reference configuration for a pinned contact line on the substrate. For all curves $f/(\gamma a)=2, \theta_p=\pi/2$, and $V_l=79\times\frac{4\pi}{3}a^3$. $(b)$ Position and depth of the free energy minimum as functions of $\theta_0$. The symbols in $(b)$ correspond to those used in $(a)$. In $(a)$ the curves end at that value of $\alpha$ at which the particle makes contact with the substrate. For pointlike particles one has $\alpha\leq\theta_0$. In all plots the lines are guides to the eye.}
	\label{theta_dependence}
\end{figure} 
For $f\neq0$ the actual contact angle $\tilde{\theta}(\phi)$ at a pinned contact line (see Eq.\ (\ref{tildetheta0})) differs from the constant value $\theta_0$ for the reference configuration ($f=0$). Up to first order in $\epsilon$ one obtains for $\theta_0=\pi/2$ and with $v(\Omega)=q\,G_B(\Omega,\Omega_1)$
\begin{equation}
\cos\tilde{\theta}(\phi)=\dfrac{a}{R_0} \dfrac{f}{\gamma a} \partial_{\theta} G_B(\Omega,\Omega_1)|_{\Omega=(\theta=\pi/2,\phi)}.
\label{tildetheta}
\end{equation}

\noindent The numerical results for $\Delta \theta(\phi):=\tilde{\theta}(\phi)-\pi/2$ and the comparison with Eq.\ (\ref{tildetheta}) are presented in Fig.~\ref{dirichlet}$(b)$. The discrepancy is typically of the order of $10\%$, which differs from the almost perfect agreement found in the case of the free energy (see Fig.~\ref{dirichlet}$(a)$).
If the particle is close to the contact line ($\alpha=72^{\circ}$) the discrepancy reaches $25\%$ which is linked to the large value of $\Delta\tilde{\theta}(\phi)\approx 30^{\circ}\approx 0.5\,[\text{rad}]$ for $\phi\approx 0$, which in turn signals that in the vicinity of the contact line and close to the particle the small gradient approximation deteriorates and the terms $O(\epsilon^2)$ in Eq.\ (\ref{tildetheta0}) become important (for the values of the parameters used in Fig.\ \ref{dirichlet}$(b)$ one has $\epsilon=0.5$).
Finally, we observe a strong dependence of the free energy on $\theta_0$ (Fig.~\ref{theta_dependence}) such that upon increasing $\theta_0$ the free energy minimum moves away from the apex but also stays away from the contact line on the substrate and its depth increases strongly if $\theta_0$ approaches $\pi$.

\subsubsection{Free contact line on the substrate}

\begin{figure}[ht]
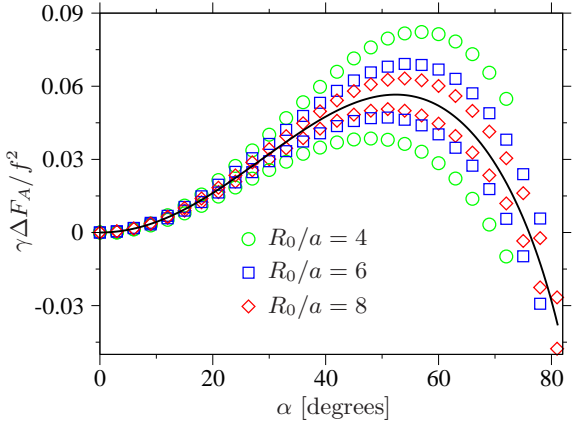

	\centering
	\psfragscanon
	\psfrag{x1}[c][c][1]{$\alpha$ [degrees]}
	\psfrag{y1}[c][c][1]{$\gamma\Delta F_A/f^2$}
	\psfrag{2}[l][l][1]{$R_0/a=2$}
	\psfrag{r4}[l][l][1]{$R_0/a=4$}
	\psfrag{r6}[l][l][1]{$R_0/a=6$}
	\psfrag{r8}[l][l][1]{$R_0/a=8$}
	\begin{overpic}[width=0.4\textwidth]{Fig14}%{c1_free_com0_8}
	\put(-5,25){\rotatebox{90}{$\gamma\Delta F_A/f^2$}}
	\put(45,-5){$\alpha$ [degrees]}	 
	\end{overpic}
	\psfragscanoff
	\caption{The rescaled excess free energy (Eq.\ (\ref{excess_free_en})) as a function of the polar angle $\alpha$ for the contact angle $\theta_0=\pi/2$ of a free contact line at the substrate, $f/(\gamma a)=1$ (above the solid line) and $f/(\gamma a)=-1$ (below the solid line), and various radii $R_0$. The solid line corresponds to the expression given by Eqs.~(\ref{greens_function}), (\ref{excess_free_enND}), and (\ref{gN}) for pointlike forces. For geometrical reasons the maximal accessible values of $\alpha$ decrease for decreasing $R_0$. For a free contact line the contact angle at the substrate remains at the value $\theta_0$ for $f\neq0$, i.e., it remains the same as for the reference configuration.}
	\label{neumann}
\end{figure}

The results for a free contact line, presented in Fig.~\ref{neumann}, demonstrate that changing the boundary conditions on the substrate can change the behavior of the particle completely. The global equilibrium position is now at the contact line but there is a deep metastable free energy minimum at the drop apex. For forces $f$ of the order of $\gamma a$ the free energy barrier is of the order of $0.05\times\gamma a^2$. Thus for micron-sized particles this is typically much higher than the free energy of thermal fluctuations, and therefore the corresponding configuration is expected to be experimentally observable. 

\begin{figure}[ht]
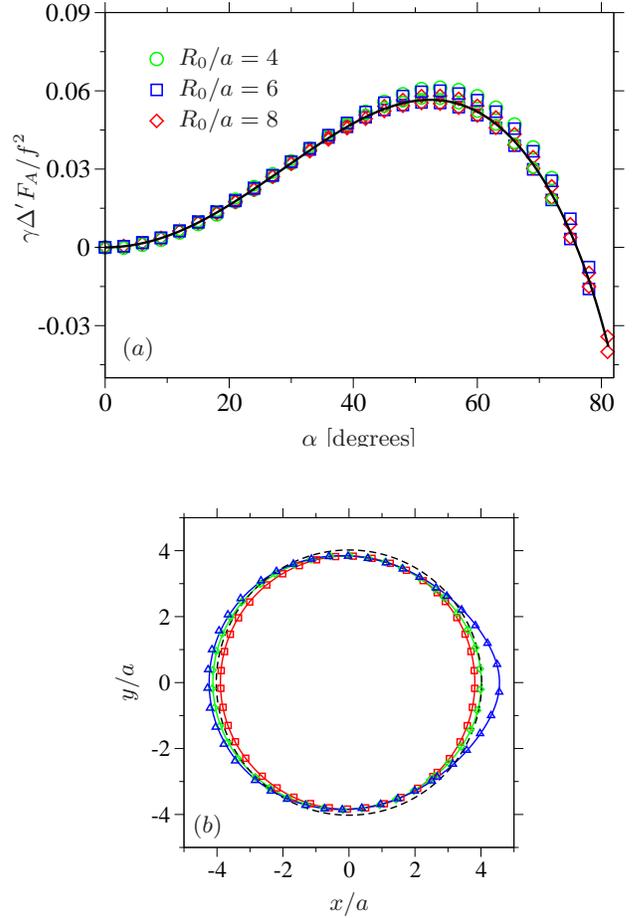

	\centering
	\psfragscanon
	\psfrag{x1}[c][c][1]{$\alpha$ [degrees]}
	\psfrag{y1}[l][l][1]{$\gamma\Delta' F_A/f^2$}
	\psfrag{r4}[l][l][1]{$R_0/a=4$}
	\psfrag{r6}[l][l][1]{$R_0/a=6$}
	\psfrag{r8}[l][l][1]{$R_0/a=8$}
	\begin{overpic}[width=0.45\textwidth]{Fig15}%{c1_free_corrected2}
	\put(17,15){$(a)$}
	\end{overpic}\\
	\psfragscanoff
	\vspace*{0.8cm}
	\psfragscanon
	\psfrag{x1}[c][c][1]{$x/a$}
	\psfrag{y1}[c][c][1]{$y/a$}
	\begin{overpic}[width=0.3\textwidth]{Fig16}%{cont_line4}
	\put(18,19){$(b)$}
	\end{overpic}
	\psfragscanoff
	\caption{$(a)$ The rescaled excess free energy (Eq.\ (\ref{excess_free_en})) for a free contact line at the substrate after taking into account the finite size correction $\delta F$ (Eq.\ (\ref{correction})) for the numerical data (symbols): $\Delta'F_A=\Delta F_A-\delta F$; the color code and the meaning of the solid line are the same as in Fig.~\ref{neumann},  $f/(\gamma a)=1$ and $f/(\gamma a)=-1$ for the points slightly above and below the solid line, respectively (visible for $\alpha>60^{\circ}$). $(b)$ The deformation of a free contact line at the substrate for $\alpha=12^{\circ}$ (squares), $\alpha=48^{\circ}$ (diamonds), and $\alpha=72^{\circ}$ (triangles), where $\theta_0=\pi/2, R_0/a=4, f/(\gamma a)=2$, and $\theta_p=\pi/2$. The dashed black line denotes the reference configuration. The solid lines are predicted analytically for pointlike forces by Eqs.\ (\ref{ND-solution}), (\ref{greens_function}), and (\ref{greens_N}). The contact angle is constant along these contact lines and equals $\theta_0=\pi/2$.}
	\label{cont_line}
\end{figure}

In the case of a free contact line we observe much larger discrepancies with the predictions of the theory for pointlike forces, revealing a dependence of $\Delta F_A$ on $f$ beyond the simple scaling $\sim f^2$. However, this deviation vanishes for increasing radii $R_0$ of the droplet indicating that this is a finite-size effect. In order to understand this effect we first consider the reference configuration, i.e., the case $f=0$. For $\theta_p=\pi/2$ the immersed part $\delta V_{ref}$ of the particle (being the intersection of the domain occupied by the particle with the spherical cap, representing the reference droplet of volume $V_l=(2\pi R_0^3/3)(1+O(a/R_0)^3)$ in the case $\theta_0=\pi/2$) has, independently of $\alpha$, the volume of $\delta V_{ref}=(2\pi a^3/3)(1+O(a/R_0))$, but the position of this cavity in the liquid depends on $\alpha$. Therefore $x_{CM,ref}$ also depends on $\alpha$ and equals (here the position of the particle is taken to have a positive $x$-component, see Fig.~\ref{sketch}):
\begin{multline}
%\begin{equation}
x_{CM,ref}(\alpha)=\dfrac{\int_{V_l}\!dV\, x}{V_l}=-\dfrac{1}{V_l}\int_{\delta V_{ref}}\!dV\, x \\
\approx -\dfrac{\delta V_{ref}}{V_l}R_0\sin\alpha\approx -\left(\dfrac{a}{R_0}\right)^2a\sin\alpha.
\label{x_cm}
%\end{equation}
\end{multline}

\noindent In contrast, in the case of a pointlike force one has $x_{CM,ref}\equiv0$. The corresponding difference in the free energy $\delta F$ can be understood as the work done by the force $f_{CM}=-f\sin\alpha$ (see the main text after Eq.\ (\ref{qcm})) applied to the center of mass in order to counterbalance the lateral component of the force $f$ (see Eq.\ (\ref{qcm})), upon displacing the center of mass from the configuration with $\alpha=0$ to $\alpha$:
\begin{multline}
%\begin{equation}
	\delta F = \int_0^{\alpha}\!d\alpha'\, f_{CM}(\alpha')d x_{CM,ref}/d\alpha'\\ =fa\left(\dfrac{a}{R_0}\right)^2\int_0^{\alpha}\!d\alpha'\,\sin\alpha'\cos\alpha'=\dfrac{fa}{2}\left(\dfrac{a}{R_0}\right)^2\sin^2\alpha.
\label{correction}
%\end{equation}
\end{multline}

\noindent This contribution~\cite{correction} should be subtracted from the numerically calculated free energy in order to facilitate the comparison with the analytical result for a pointlike force. It is linear in $f$, which explains the aforementioned deviation from the scaling $\sim f^2$. For large drops it vanishes $\sim(a/R_0)^2$ so that for $(a/R_0)\lesssim0.1$ it can practically be neglected (compare Fig.~\ref{neumann}). However, for smaller droplets this correction has to be taken into account in order to obtain agreement with the perturbation theory (see Fig.~\ref{cont_line}$(a)$).

%%%%%%%%%%%%%%%%%%%%%%%%%%%%%%%%%%%%%%%%%%%%%%%%%%%%%%%%%%%%%%%%%%%%%%%%%%%%%%%%%%%%%
%%%%%%%%%%%%%%%%%%%%%%%%%%%%%%%%%%%%%%%%%%%%%%%%%%%%%%%%%%%%%%%%%%%%%%%%%%%%%%%%%%%%%

\section{Summary and conclusions}

We have studied the influence of curvature and confinement on the free energy of a particle of radius $a$ floating at the surface of a sessile droplet of radius $R_0$ and exposed to external forces $f$ acting in the direction normal to the unperturbed droplet surface (see Fig.\ \ref{sketch}). Our results can be summarized as follows. 

If the particle is at the drop apex the system exhibits axial symmetry which allows one to obtain exact analytic solutions for the droplet shape and expressions for the surface free energy depending on the immersion of the particle into the liquid phase (see Fig.\ \ref{symmetric}). The comparison of these analytic results with those obtained in Appendix C within linear perturbation theory indicates that non-linear effects contained in the full Young-Laplace equation (see Eq.\ (\ref{young-laplace})) can be neglected even for relatively large particles (see Fig.\ \ref{symmetric_analytic}). The comparison of these full analytic results with those obtained by a numerical minimization of the free energy (Sec.\ V) validates the high accuracy of the numerical code (Fig.\ \ref{symmetric}).

In the cases without axial symmetry the condition of balance of forces acting on the droplet in lateral directions requires either a fixed lateral position of the center of mass of the droplet (leaving the contact line free: model $A$) or a pinned contact line at the substrate (leaving the center of mass free: model $B$). Using a perturbation theory for small deformations of the droplet we have derived a free energy functional appropriate for both models. In the case of a free contact line at the substrate, the ensuing changes in the substrate-liquid surface energy have been also taken into account (Eq.\ (\ref{functional})). In model $A$ the contact angle at the substrate maintains its value $\theta_0$ of the unperturbed droplet whereas in model $B$ it varies along the contact line and can differ from $\theta_0$. The pulling of the particle by an external force $f$ has been implemented by introducing an effective pressure field $\pi(\Omega)$ (see Fig.\ \ref{virtual} and Eq.\ (\ref{exppi})), which enters the linear Young-Laplace equation determining the small deformations of the droplet (Eq.\ (\ref{helmholtz})). Similarly, the fixing of the center of mass is implemented by an effective pressure field $\pi_{CM}(\Omega)$ (see Eqs.\ (\ref{helmholtz}) and (\ref{piCM})).
We have shown that in the limit of small particles, i.e., for $a/R_0\rightarrow0$, the free energy of the sessile droplet (Eq.\ (\ref{F_ND})), expressed in terms of a Green's function satisfying the boundary conditions at the substrate corresponding to either a free or a pinned contact line (Eqs.\ (\ref{greens_N}) and (\ref{greens_D})), does not depend on the size of the particle but only on the pulling force $f$ (Eq.\ (\ref{pressure3})), the contact angle $\theta_0$ of the unperturbed droplet at the substrate, and on the angular position of the particle $\alpha$ (Fig.\ \ref{sketch}). 

If the contact angle $\theta_0$ at the substrate equals $\pi/2$ one can exploit an analogue of the method of images known from electrostatics in order to calculate the surface free energy (in excess over the surface free energy of the reference configuration of a drop shape given by a spherical cap) depending on $\alpha$. The boundary conditions at the substrate are met by introducing an image particle at the virtual droplet hemisphere below the substrate surface. In analogy to electrostatics a pinned contact line corresponds to Dirichlet boundary conditions (see Fig.\ \ref{imageD}) and a free contact line with fixed contact angle corresponds to Neumann boundary conditions (see Fig.\ \ref{imageN}). A change of the type of boundary conditions leads to a change of sign of the capillary charge associated with the image particle; accordingly, the excess free energy, which is proportional to the interaction energy of the original particle with its image, also changes sign. Further analysis shows that due to the conditions of force balance and volume constraint the Green's function requires additional terms (see Eq.~(\ref{greens_D})), but they do not change the results qualitatively. 

If one treats the particle as a capillary monopole which exerts a pointlike force on the fluid interface, the excess free energy can be expressed in terms of the Green's function $G(\bar{\theta})$ (Eq.~(\ref{greens_function})) derived by Morse and Witten~\cite{Morse1993} for a free spherical drop, where $\bar{\theta}$ is the angle between the direction of observation of the interface displacement and the direction into which the pointlike external force is applied, both in radial direction from the center of the reference configuration, i.e., normal to the surface of the unperturbed droplet (see Fig.\ \ref{sketch}). For $\bar{\theta}\ll 1$ the Green's function exhibits a characteristic logarithmic divergence recovering the known corresponding behavior for the case of a flat interface; the size of the particle acts as a natural cut-off. Compared to this latter case the Green's function of the finite-sized drop gives rise to a new interesting effect due to the non-monotonic behavior of $G(\bar{\theta})$ for intermediate angles. As a consequence, besides the known phenomena of attraction of the particle to the free contact line and repulsion from the pinned contact line, we have found a deep local free energy minimum for the particle being at the drop apex for a free contact line (Fig.~\ref{neumann}) and a global free energy minimum at an intermediate angle $\alpha$ (Fig.~\ref{sketch}) for a pinned contact line (Fig.~\ref{dirichlet}). The positions of these free energy minima are de facto independent of the strength $f$ of the external force acting on the particle, the droplet radius $R_0$, and the contact angle $\theta_p$ at the particle, while their depths are proportional to $f^2$. For a particle of radius $1 \mu m$ and an interface with surface tension $\gamma=0.05 N m^{-1}$ the depth is of the order of $10^5 k_B T$ and therefore the corresponding configuration is expected to be experimentally observable. A pulling force $f$ can be applied to the particle by using optical tweezers (see, for example, Ref.~\cite{Blickle2009}). This technique would also enable one to measure the force tangential to the liquid-gas interface corresponding to the derivative $d(\Delta F_{\sigma})/d(R_0\alpha)$ of the free energy ($\sigma=A,B$) with respect to the lateral displacement of the particle along the interface.

%\noindent (iv) 
Moreover, our numerical results in Fig.~\ref{theta_dependence} show that for model $B$ the depth of the free energy minimum increases rapidly upon increasing the contact angle $\theta_0$ at the substrate towards $180^{\circ}$. In the case of model $A$, the shape of the contact line at the substrate as obtained from numerical calculations agrees very well with the predictions of the analytic expressions given in Eqs.\ (\ref{greens_N}) and (\ref{greens_function}) (see Fig.\ \ref{cont_line}$(b)$), whereas the free energy (Fig.\ \ref{neumann}), due to the condition of fixed center of mass, is more sensitive to the finite size $a$ of the particle and converges to the expression for a pointlike particle for $a/R_0\lesssim 0.1$. On the other hand, in the case of a pinned contact line the finite-size effects are negligible (Fig.\ \ref{dirichlet}$(a)$). In this case the contact angle differs from its reference value $\theta_0$ and varies along the contact line (Fig.\ \ref{dirichlet}$(b)$), in agreement with the analytic approximation (Eqs.\ (\ref{tildetheta0}) and (\ref{tildetheta}) for $\theta_0=\pi/2$).

The overall very good agreement with the pointlike force approximation, even for relatively large particles such as with $a/R_0=0.25$, indicates that higher order corrections in $a/R_0$ are very small. Indeed, in the case of a spherical particle the capillary dipole must vanish due to the vanishing torque on the particle and the first subleading term stems from the interaction energy of a monopole with an induced quadrupole yielding a correction of the order $(a/R_0)^4$.

\acknowledgements
We gratefully acknowledge fruitfull discussions with Alvaro Dom\'{\i}nguez, Martin Oettel, and Nelson Bernardino. 
One of the authors (J.G.) thanks  Ken Brakke for tremendous support and patience in providing tips and source codes concerning the software Surface Evolver, Alvaro Dom\'{\i}nguez for encouragement and Fabian D{\"o}rfler for discussions.
%One of the authors (J.G.) thanks  Ken Brakke for tremendous support and patience in providing tips and source codes concerning the software Surface Evolver, and Fabian D{\"o}rfler for discussions.

\appendix

%%%%%%%%%%%%%%%%%%%%%%%%%%%%%%%%%%%%%%%%%%%%%%%%%%%%%%%%%%%%%%%%%%%%%%%%%%%%%%%%%%%%%
%%%%%%%%%%%%%%%%%%%%%%%%%%%%%%%%%%%%%%%%%%%%%%%%%%%%%%%%%%%%%%%%%%%%%%%%%%%%%%%%%%%%%

\section{Free energy of a sessile droplet}
In this appendix we derive the expression for the free energy of a sessile droplet in mechanical equilibrium and subjected to a pressure field $\pi(\Omega)$ (Eq.\ (\ref{F_ND})). In terms of perturbation theory, the free energy both for model $A$ and model $B$ is given by the functional in Eq.\ (\ref{small_def2}) evaluated for $v$ obeying the Young-Laplace equation (\ref{helmholtz}) with $\pi_{CM}=0$ for model $B$ and with the condition in Eq.\ (\ref{xcmu2}) for model $A$. This yields
\begin{multline}
	F =\dfrac{f^2}{\gamma} \int_{\Omega_0}\! d\Omega\, \left[\frac{1}{2}\nabla_a\big(v\nabla_a v\big)-\frac{1}{2}v\big(\nabla_a^2 v+ 2v\big)\right.\\
	 \left. -\big(\pi(\Omega)+\mu\big)v \right] -\dfrac{f^2}{2\gamma}\cos\theta_0\int_{0}^{2\pi}\! d\phi\,\big(v\vert_{\theta_0}\big)^2=\\
	 =-\dfrac{f^2}{2\gamma} \int_{\Omega_0}\! d\Omega\, \pi(\Omega)v\\
	 + \dfrac{f^2}{2\gamma} \int_{0}^{2\pi}\! d\phi\, v\vert_{\theta_0}\big(\sin\theta_0\partial_{\theta} v\vert_{\theta_0}-\cos\theta_0 v\vert_{\theta_0}\big),
	\label{derivation1}
\end{multline}

\noindent where the second equality follows from applying Gau{\ss}' theorem, Eq.\ (\ref{helmholtz}), Eq.\ (\ref{xcmu2}), and the constant volume constraint (Eq.\ (\ref{volume2})). The second term in the last expression vanishes for boundary conditions corresponding to either a free or a pinned contact line at the substrate (see Eqs.\ (\ref{Rob}) and (\ref{Dir})). This leads to the free energy in the form in Eq.\ (\ref{F_ND}).

%%%%%%%%%%%%%%%%%%%%%%%%%%%%%%%%%%%%%%%%%%%%%%%%%%%%%%%%%%%%%%%%%%%%%%%%%%%%%%%%%%%%%
%%%%%%%%%%%%%%%%%%%%%%%%%%%%%%%%%%%%%%%%%%%%%%%%%%%%%%%%%%%%%%%%%%%%%%%%%%%%%%%%%%%%%

\section{Exact results for axially symmetric configurations with a free contact line at the substrate}

\begin{figure}[ht]
	\centering
	\psfragscanon
	\psfrag{o}[c][c][1]{$0$}
	\psfrag{r}[c][c][1]{{\large$r$}}
	\psfrag{z}[c][c][1]{{\large$z$}}
	\psfrag{r0}[c][c][1]{$R_0\sin\theta_0$}
	\psfrag{t0}[c][c][1]{$\theta_0$}
	\psfrag{z0}[c][c][1]{$z_0$}
	\psfrag{z1}[c][c][1]{$z_0-1$}
	\psfrag{tp}[c][c][1]{$\theta_p$}
	\psfrag{b0}[c][c][1]{$\beta_0$}
	\psfrag{su}[c][c][1]{substrate}
	\psfrag{li}[c][c][1]{liquid}
	\psfrag{ga}[c][c][1]{gas}
	\psfrag{zr}[c][c][1]{$\quad z(r)$}
	\psfrag{nic}[c][c][1]{}
	\begin{overpic}[width=0.35\textwidth]{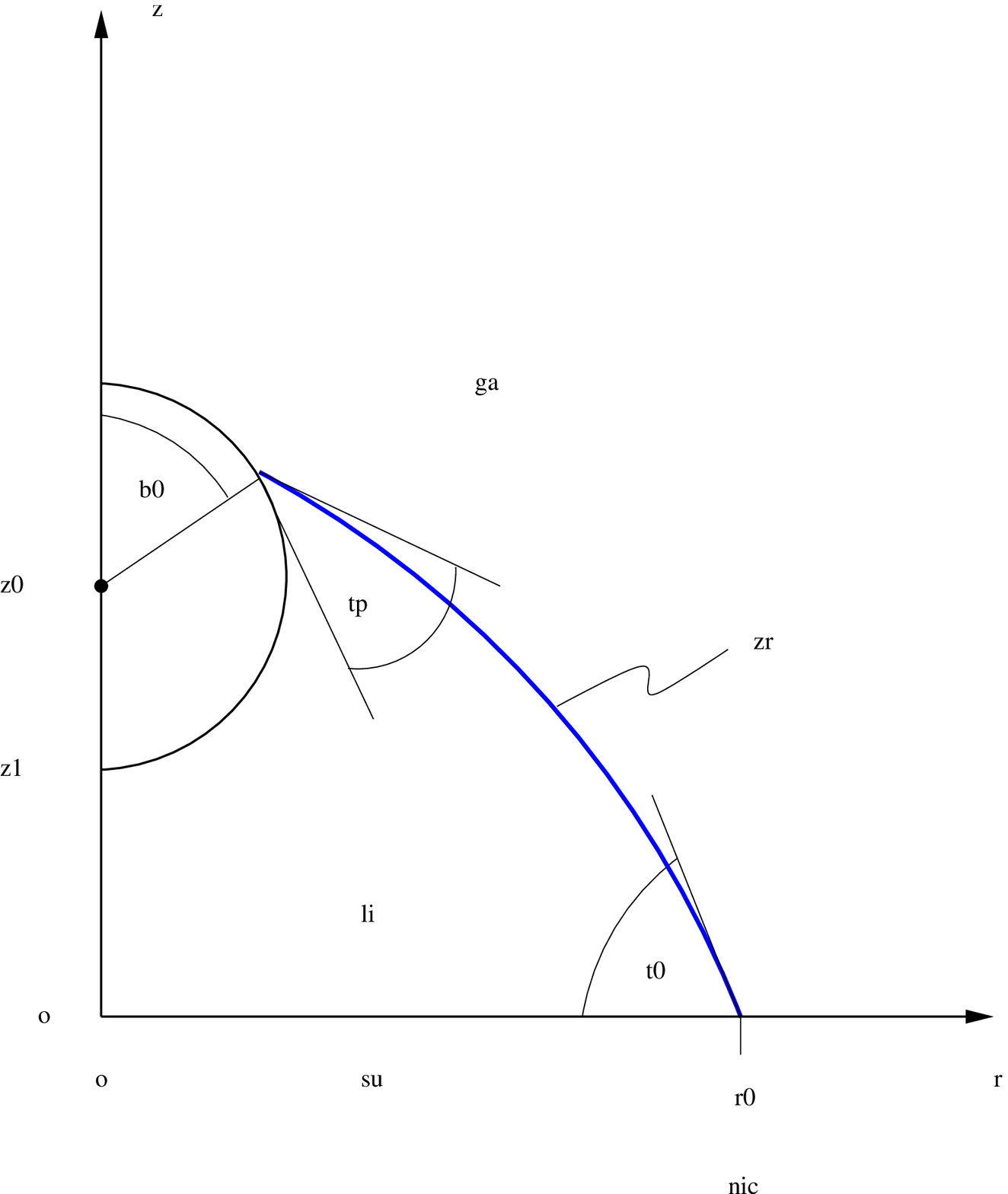}%{reference_conf4}
	\put(60,90){{\large$(a)$}}
	\end{overpic}
	\psfragscanoff\\
 	\psfragscanon
	\psfrag{o}[c][c][1]{$0$}
	\psfrag{r}[c][c][1]{{\large$r$}}
	\psfrag{z}[c][c][1]{{\large$z$}}
	\psfrag{zh}[c][c][1]{$z_0+h\;$}
	\psfrag{r0}[c][c][1]{$r_0$}
	\psfrag{r1}[r][r][1]{$r_1$}
	\psfrag{rm}[c][c][1]{$r_m$}
	\psfrag{t0}[c][c][1]{$\theta_0$}
	\psfrag{tp}[c][c][1]{$\theta_p$}
	\psfrag{b}[c][c][1]{$\beta$}
	\psfrag{btp}[l][l][1]{$\beta-\theta_p$}
	\psfrag{psi}[c][c][1]{$\psi (r)$}
	\psfrag{sin}[l][l][1]{$\sin\beta$}
	\begin{overpic}[width=0.35\textwidth]{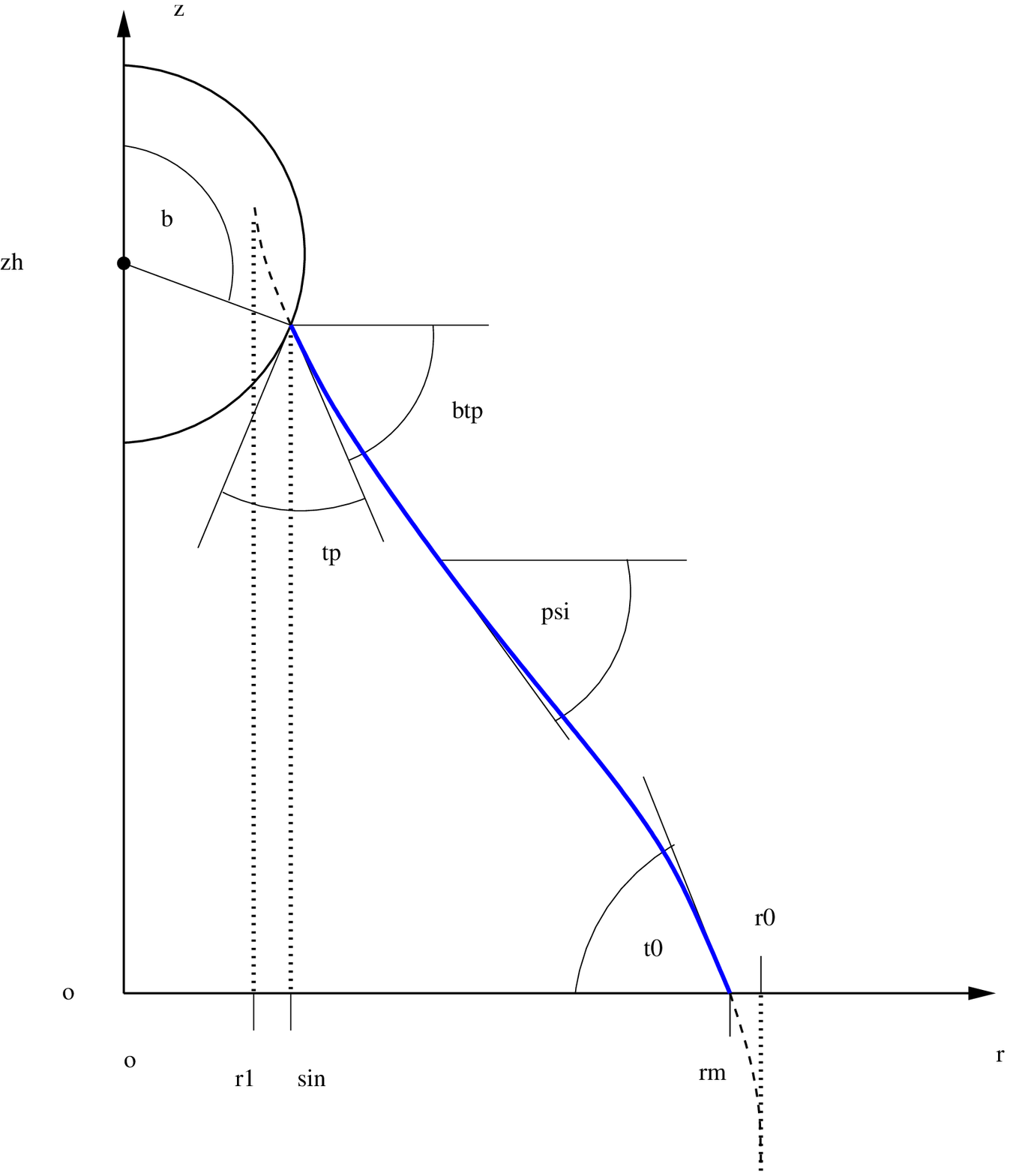}%{axisymmetric8}
	\put(60,90){\large{$(b)$}}
	\end{overpic}
	\psfragscanoff 
	\caption{Cross section of an axially symmetric configuration of the droplet and the particle; $(a)$ reference configuration for $f=0$, which leads to a liquid-gas interface with the shape of a spherical cap; $(b)$ sketch of a configuration for $\beta\neq\beta_0$ corresponding to $f\neq 0$. This sketch corresponds to model $A$ for which $\theta_0$ is kept fixed and $r_m\neq R_0\sin\theta_0$. The dashed lines denote continuations of the analytical profile corresponding to an unduloid or a nodoid beyond the physical regime bounded by $r=\sin\beta$ and $r=r_m$ and terminating at maximal and minimal radii $r_0$ and $r_1$, respectively, with $\dot{z}(r_0)=\dot{z}(r_1)=-\infty$. Note that even for $h>1$ this configuration can be stable (compare Fig.\ \ref{symmetric}$(a)$).}
	\label{axisymmetric}
\end{figure}

In this appendix we study the shape and the free energy of the droplet with the adsorbed particle at its apex, resulting in an axially symmetric configuration. In the following calculations we take $a$ as the unit of length and $\gamma a^2$ as the unit of energy. For simplicity we assume that the interface has no overhangs ($\theta_0\leq\pi/2$) so that the shape of the droplet can be described by $z=z(r)$. We also assume a free contact line and a given contact angle $\theta_0$ at the substrate (see Fig.~\ref{axisymmetric}). A similar analysis can be performed also for a pinned contact line, in which case the surface energy associated with the substrate would be constant. The problem with a free contact line is actually analytically more involved, because in this case the surface free energy associated with the substrate can vary. Therefore it is particularly interesting and instructive to incorporate the contribution from the liquid-substrate interface into the total free energy.  Our goal is to calculate the free energy $\tilde{F}_{\sigma0}$, introduced implicitly as the inverse Legendre transform of $F_{A0}$ in Eq.\ (\ref{free_en_lagrange}) (see also Eqs.\ (\ref{excess_free_en}) and (\ref{functional})) with $\sigma=A$ corresponding to a free contact line, as a function of the vertical position $h$ of the particle and the independent system parameters $\theta_0$, $\theta_p$, $R_0$ (the latter expressed in units of $a$). To this end we first consider the free energy (compare the analogous expression in Eq.\ (\ref{minimizeN}), noting that here there is no force $f_{CM}$ needed):
\begin{equation}
	\tilde{F}_{A0}^{\star}(h,\theta_0,\theta_p,R_0,\lambda) = \min_{\{z(r)\}}\tilde{\mathcal{F}}[\{z(r)\}],
	\label{Fh}
\end{equation}

\noindent with the free energy functional $\tilde{\mathcal{F}}$ as
\begin{multline}
%\begin{equation}
	\tilde{\mathcal{F}}[\{z(r)\}; h,\theta_0,\theta_p,V_l,\lambda] = S_{lg}-S_{lg,ref}\\
	 - (S_{0l}-S_{0l,ref})\cos\theta_0 - (S_{pl}-S_{pl,ref})\cos\theta_p - \lambda(V-V_l),
	\label{functional0}
%\end{equation}
\end{multline}

\noindent where the volume $V_l$ of liquid is a function of the contact angles $\theta_0$ and $\theta_p$, as well as of the droplet radius $R_0$ in the reference configuration, i.e., $V_l=V_l(\theta_0,\theta_p,R_0)$. The Lagrange multiplier $\lambda$ can be determined from the condition $V=V_l$, which renders $\lambda=\lambda(h,\theta_0,\theta_p,R_0)$ and which upon insertion into $\tilde{F}_{A0}^{\star}(h,\theta_0,\theta_p,R_0,\lambda)$ yields the desired free energy $\tilde{F}_{A0}(h,\theta_0,\theta_p,R_0)$.

It is convenient to express the contact areas $S_{0l}$, $S_{pl}$, and $S_{lg}$ in terms of two auxiliary variables: the angle $\beta $ describing the position of the contact line at the particle and the radius $r_m$ of the circle formed by the contact line at the substrate (see Fig.\ \ref{axisymmetric}$(b)$). In terms of these variables one has 
\begin{align}
	&S_{0l} = \pi r_m^2, \label{S0l}\\
	&S_{pl} = 2\pi (1+\cos\beta)=4\pi\cos^2(\beta/2), \label{Spl}\\
	&S_{lg} = 2\pi\int_{\sin\beta}^{r_m}\! dr\, r \sqrt{1+\dot{z}^2},
	\label{area}
\end{align} 

\noindent with $\dot{z}=dz(r)/dr$. The volume of liquid $V$ can be expressed as
\begin{align}
	&V =  V_{lg} + \pi(\sin^2\beta)z(r=\sin\beta)- \dfrac{4\pi}{3}f_0(\pi-\beta)
	\label{volume}
\end{align} 

\noindent with $f_0(x)$ given by Eq.\ (\ref{f0}) and
\begin{align}
	&V_{lg} = 2\pi\int_{\sin\beta}^{r_m}\! dr\, r z(r),
	\label{volume0}
\end{align} 

\noindent where $V_{lg}$ is the volume enclosed between the substrate and the liquid-gas interface, the second term in Eq.\ (\ref{volume}) is the volume between the substrate and the circular disc determined by the contact line at the particle, and the third term is the volume of that part of the particle which is immersed in the liquid. 

The equilibrium profile $z(r)$, which minimizes $\tilde{\mathcal{F}}$ obeys the Young-Laplace equation in cylindrical coordinates (see, e.g., Ref.~\cite{Langbein_book}):
\begin{equation}
 	\dfrac{1}{r}\dfrac{d}{dr}\dfrac{r\dot{z}}{\sqrt{1+\dot{z}^2}} = -\lambda.
	\label{young-laplace}
\end{equation}

\noindent The boundary conditions are determined by Young's law at the particle and at the substrate which can be shown to follow from the condition of vanishing of the variation of $\tilde{\mathcal{F}}$ at the boundaries:
\begin{align}
	&\psi(r=\sin\beta)=\beta-\theta_p, \label{young1} \\
	&\psi(r=r_m)=\theta_0 \label{young2},
\end{align} 

\noindent where $\psi=\psi(r)$ is the angle between the $r$-axis and the tangent of the profile $z(r)$ (see Fig.~\ref{axisymmetric}$(b)$) defined by
\begin{equation}
	\sin\psi(r):=-\dfrac{\dot{z}}{\sqrt{1+\dot{z}^2}}.
\end{equation}

\noindent The first integral of Eq.\ (\ref{young-laplace}) reads
\begin{equation}
	\dfrac{\dot{z}}{\sqrt{1+\dot{z}^2}}\equiv-\sin\psi(r) = -\dfrac{\lambda r}{2}+\dfrac{c}{r},
	\label{first_integral}
\end{equation}

\noindent  where $c$ is an integration constant. When evaluated at the boundaries and using Eqs.\ (\ref{young1}) and (\ref{young2}) this leads to the following set of equations:
\begin{align}
	&-\sin(\beta-\theta_p) = -\dfrac{\lambda\sin\beta}{2} + \dfrac{c}{\sin\beta},\\
	&-\sin\theta_0 = -\dfrac{\lambda r_m}{2} + \dfrac{c}{r_m}.
\end{align} 
 
\noindent Solving with respect to $\lambda$ and $c$ one obtains
\begin{align}
	&\lambda=\dfrac{2[r_m\sin\theta_0-\sin\beta\sin(\beta-\theta_p)]}{r_m^2-\sin^2\beta}=\lambda(r_m,\beta), \label{lagrange}\\
	&c=-\dfrac{r_m^2\sin\beta\sin(\beta-\theta_p)-r_m\sin\theta_0\sin^2\beta}{r_m^2-\sin^2\beta}=c(r_m,\beta) \label{const}.
\end{align} 

\noindent The equilibrium profile $z(r)$ is obtained by solving Eq.\ (\ref{first_integral}) for $\dot{z}$ and by subsequently integrating:
\begin{multline}
	z(r) = -\dfrac{1}{\lambda}\int_{\sin\beta}^{r_m}\! dr \dfrac{\lambda r^2-2c}{\sqrt{(r^2-r_1^2)(r_0^2-r^2)}}= \\
	= -r_0\big[E(\phi,q) - E(\phi_2,q)\big]+\dfrac{2c}{\lambda r_0}\big[F(\phi,q) - F(\phi_2,q)\big].
	\label{z_profile}
\end{multline}

\noindent The integral in Eq.\ (\ref{area}) can be evaluated as
\begin{multline}
	S_{lg}=\dfrac{4\pi}{|\lambda|}\int_{\sin\beta}^{r_m}\! dr \dfrac{r^2}{\sqrt{(r^2-r_1^2)(r_0^2-r^2)}}=\\
	=4\pi\dfrac{r_0}{|\lambda|}\bigg(E(q)-E(\phi_1,q)-E(\phi_2,q)\\
	+\dfrac{1}{r_0\sin\beta}\sqrt{(r_0^2-\sin^2\beta)(\sin^2\beta-r_1^2)}\bigg).
	\label{Slg}
\end{multline}

\noindent The expression in Eq.\ (\ref{volume0}) can be evaluated by integrating by parts, which leads to
\begin{multline}
	V_{lg} = \dfrac{\pi}{\lambda}\int_{\sin\beta}^{r_m}\! dr \dfrac{\lambda r^4-2cr^2}{\sqrt{(r^2-r_1^2)(r_0^2-r^2)}}-\pi(\sin^2\beta) z(r=\sin\beta)\\
	=\pi r_0\Bigg( \kappa \big[E(q) - E(\phi_1,q) - E(\phi_2,q)\big]\\
	-\dfrac{r_1^2}{3}\big[K(q) - F(\phi_1,q) - F(\phi_2,q)\big]\\
	+\left[\dfrac{\sin\beta}{3}+\dfrac{\kappa}{\sin\beta}\right]
	\sqrt{(r_0^2-\sin^2\beta)(\sin^2\beta-r_1^2)}\\
	-\dfrac{r_m}{3}\sqrt{(r_0^2-r_m^2)(r_m^2-r_1^2)} \Bigg)
	-\pi(\sin^2\beta) z(r=\sin\beta),
\label{Vlg}
\end{multline}

\noindent where $F(\phi,k)$ and $E(\phi,k)$ are the incomplete elliptic integrals of the first and second kind, respectively, whereas $K(k)$ and $E(k)$ are the corresponding complete elliptic integrals (see Ref.~\cite{Prudnikov_book1}). Moreover, one has
\begin{align}
	&\kappa =     \dfrac{2c}{3\lambda}+\dfrac{8}{3\lambda^2},\label{kappa}\\
	&q =          \sqrt{1-\dfrac{r_1^2}{r_0^2}},\\
	&\sin\phi_1 = \dfrac{r_0}{\sin\beta}\sqrt{\dfrac{\sin^2\beta-r_1^2}{r_0^2-r_1^2}},
\end{align}

\noindent and
\begin{align}
	&\sin\phi_2 = \sqrt{\dfrac{r_0^2-r_m^2}{r_0^2-r_1^2}}.
\end{align} 

\noindent The shape of the droplet is always a section of a nodoid or an unduloid~\cite{Langbein_book} characterized by a maximal and a minimal radius $r_0$ and $r_1$ (see Fig.\ \ref{axisymmetric}$(b)$), respectively, which are two distinct solutions of the equation $\dot{z}(r)=-\infty$ and are given by \begin{align}
	&r_0 = \dfrac{1}{\lambda}\left(\sqrt{1+2\lambda c}+1\right),\\
	&r_1 = \dfrac{1}{\lambda}\left(\sqrt{1+2\lambda c}-1\right).
	\label{radius1}
\end{align} 

\begin{figure}[ht]
	\centering
	\psfragscanon
	\psfrag{r0}[l][l][1.2]{$R_0$}
	\psfrag{a}[l][l][1.2]{$a$}
	\psfrag{b0}[l][l][1.2]{$\beta_0$}
	\psfrag{tp}[l][l][1.2]{$\theta_p$}
	\psfrag{t0}[l][l][1.2]{$\theta_0$}
	\psfrag{d}[l][l][1.2]{$\delta$}
	\psfrag{z0}[l][l][1.2]{$z_0$}
	\psfrag{cl}[c][c][1.2]{CL}
	\includegraphics[width=0.45\textwidth]{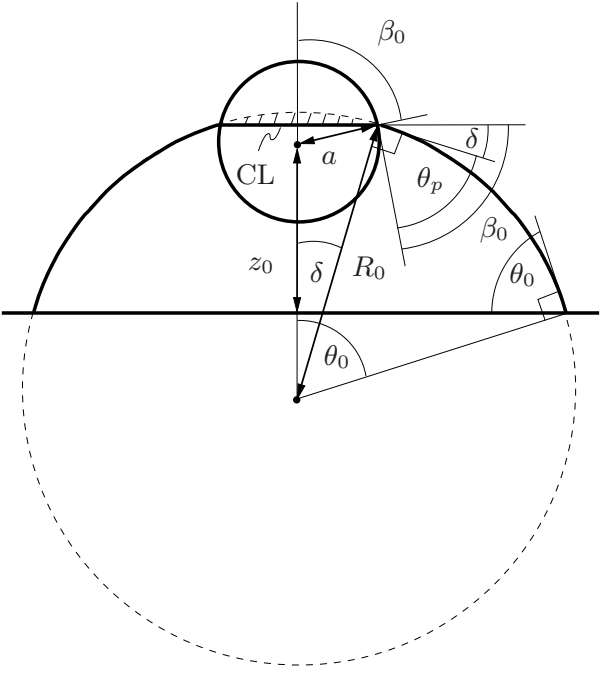}%{variables4}
	\caption{Geometry of the reference configuration. The angle $\beta_0$ can be found by comparing two expressions for the radius of the contact line at the particle: $a\sin\beta_0=R_0\sin\delta$, where $\delta$ can be found to be equal to $\beta_0-\theta_p$ (which then yields Eq.\ (\ref{beta0})). For calculating the volume $V_l$ of the liquid one must subtract from the volume of the spherical cap of radius $R_0$ the volume of the immersed part of the particle (beneath the contact line CL) and in addition the small spherical cap indicated as a hatched region (see Eq.\ (\ref{volume_ref})).}
	\label{reference}
\end{figure}

So far the interface profile (Eq.\ (\ref{z_profile})) and the free energy (Eqs.\ (\ref{Fh})-(\ref{Spl}) and (\ref{Slg})) are determined in terms of the independent variables $r_m$ and $\beta$ (through Eqs.\ (\ref{lagrange}), (\ref{const}) and (\ref{kappa})-(\ref{radius1})). One can replace them by the physically more directly accessible variables $h$ and $V_l$. To this end we note that the vertical displacement $h$ of the particle is determined by (see Fig.~\ref{axisymmetric}$(b)$)
\begin{equation}
	z_0+h = z(r=\sin\beta)-\cos\beta,
	\label{displacement}
\end{equation}

\noindent where $z_0=z_0(\theta_0,\theta_p,R_0)$ is the height of the particle center above the substrate in the reference configuration with the droplet shape given by the cap of a sphere (see, c.f., Eq.\ (\ref{implicit2}) and Fig.\ \ref{reference}). This renders the set of equations
\begin{align}
	&h=h(r_m,\beta,V_l),\label{implicit1}\\
	&V_l=V(r_m,\beta),\label{Vrbeta}
\end{align} 

\noindent where $h(r_m,\beta,V_l)$ is given by Eq.\ (\ref{displacement}) (with the dependence on $r_m$ entering via $z(r)$ and the dependence on $V_l$ entering via $z_0(R_0)$ and Eq.\ (\ref{volume_ref})) and $V(r_m,\beta)$ is defined (suppressing the explicit dependence on $\theta_0$ and $\theta_p$) via Eqs.\ (\ref{volume}) and (\ref{Vlg}) together with (\ref{lagrange}), (\ref{const}) and (\ref{kappa})-(\ref{radius1}). Equations (\ref{implicit1}) and (\ref{Vrbeta}) provide implicitly the maximal radius $r_m$ and the angle $\beta$ as functions of $V_l$ and $h$. 

Finally, with keeping in mind that here all length scales are measured in units of $a$, the constant liquid volume $V_l$ can be expressed in terms of the materials parameters $\theta_0$ and $\theta_p$ as well as the drop size $R_0$. As one can infer from the geometrical features shown in Fig.~\ref{reference}, $V_l$ is given by
\begin{multline}
%\begin{equation}
 V_l(R_0,\theta_0,\theta_p) = \dfrac{4\pi}{3}\Big(\big[f_0(\theta_0)-f_0(\beta_0-\theta_p)\big]R_0^3\\
 -f_0(\pi-\beta_0)\Big), \label{volume_ref}
%\end{equation}
\end{multline}

\noindent with
\begin{equation}
 \sin\beta_0 = R_0\sin(\beta_0-\theta_p), \label{beta0}
\end{equation}

\noindent so that 
\begin{equation}
 \beta_0(R_0,\theta_p) = \arcsin\left((R_0\sin\theta_p)/(R_0^2-2R_0\cos\theta_p+1)^{1/2}\right),
\end{equation}

\noindent and with the geometric factor $f_0(x)$ given by Eq.\ (\ref{f0}). Additionally, since the presence of the particle does not change the contact angle $\theta_0$, the liquid volume $V_l$ can be expressed in terms of the size $\bar{R}_0=[3V_l/(4\pi f_0(\theta_0))]^{1/3}$ of the droplet without the particle. The vertical position $z_0$ of the particle in the reference configuration can be expressed as
\begin{equation}
	z_0(R_0,\theta_0,\theta_p) = R_0\cos(\beta_0-\theta_p)-R_0\cos\theta_0-\cos\beta_0.
	\label{implicit2}
\end{equation}

\noindent The sets of equations (\ref{implicit1}) - (\ref{implicit2}) (which must be solved numerically) implicitly yield $\beta=\beta(h,R_0)$ and $r_m=r_m(h,R_0)$, which can be substituted into the expressions for the surface areas in Eqs.\ (\ref{S0l}), (\ref{Spl}), and (\ref{Slg}). Finally, the free energy functional $\tilde{\mathcal{F}}$ given as a combination of these areas (Eq.\ (\ref{functional0})) yields the free energy $\tilde{F}_{A0}(h,\theta_0,\theta_p,R_0)$. The results of these calculations for $\theta_0=\pi/3,\theta_p=\pi/2$, and $V_l=79\times(4\pi/3)$ are presented in Figs.\ \ref{symmetric} and \ \ref{symmetric_analytic}. 

%%%%%%%%%%%%%%%%%%%%%%%%%%%%%%%%%%%%%%%%%%%%%%%%%%%%%%%%%%%%%%%%%%%%%%%%%%%%%%%%%%%%%
%%%%%%%%%%%%%%%%%%%%%%%%%%%%%%%%%%%%%%%%%%%%%%%%%%%%%%%%%%%%%%%%%%%%%%%%%%%%%%%%%%%%%

\section{Droplet shapes from perturbation theory for configurations with axial symmetry and arbitrary $\theta_0$}
A modification of the method of images can be used for arbitrary $\theta_0$ if the point force $f=q\epsilon\gamma R_0$ is placed at the apex of the sessile droplet. The virtual reference droplet (for $f=0$) is constructed as a smooth continuation of the actual droplet such that it completes the full sphere and the image of amplitude $f'=f$ is placed at the ''south pole'' $\theta'=\pi$. In this configuration forces acting on the union of the real and the virtual parts of the droplet are balanced. We exploit the freedom to add a constant and a term proportional to $\cos\theta$, such that the axially symmetric solution
\begin{equation}
 v_0(\theta)=q[G(\theta)+G(\pi-\theta)+H_0\cos\theta+I_0]
\label{v_symmetric}
\end{equation}

\noindent conserves the volume and fulfills the given boundary conditions. Moreover, this solution fulfills Eq.\ (\ref{helmholtz}) with $\pi(\Omega)=q\,\delta(\Omega)$ and $\pi_{CM}(\Omega)=0$ upon choosing $I_0=-\mu/2-1/(4\pi)$. The value of $\mu$, given by Eqs.\ (\ref{musigmaA}) and (\ref{musigmaB}) for $\alpha=0$, depends on the boundary condition at the substrate. In order to be able to make a comparison with the nonlinear theory in Appendix B here we take the contact line at the substrate to be free. (For the case of a pinned contact line see Appendix B in Ref.~\cite{Oettel2005}.) Using the expression in Eq.\ (\ref{musigmaA}) for $\alpha=0$ yields
\begin{equation}
 I_0=\dfrac{\cos\theta_0(1+\cos\theta_0)}{4\pi(2+\cos\theta_0)(1-\cos\theta_0)}.
\label{P0}
\end{equation}

\noindent Inserting Eqs.\ (\ref{v_symmetric}) and (\ref{P0}) into the volume constraint $\int_0^{\theta_0}\!d\theta\,\sin\theta v_0(\theta)=0$ one obtains 
\begin{equation}
 H_0=\dfrac{1}{2\pi}\left[\ln\tan(\theta_0/2)-\dfrac{\cos\theta_0}{(2+\cos\theta_0)(1-\cos\theta_0)}\right].
\label{Q0}
\end{equation}

\noindent Due to $I_0(\theta_0=\pi/2)=0=H_0(\theta_0=\pi/2)$, for $\theta_0=\pi/2$ the solution in Eq.\ (\ref{v_symmetric}) reduces to 
\begin{equation}
 v_0(\theta;\theta_0=\pi/2)= q[G(\theta)+G(\pi-\theta)]\equiv q\,G_A(\Omega,\Omega_1=0),
\end{equation}

\noindent which reproduces the result of the perturbation theory for model $A$ (Eq.\ (\ref{greens_N})) for $\alpha=0$. In order to express the deformation $u$ in the dimension of length the solution in Eq.\ (\ref{v_symmetric}) has to be multiplied by $R_0\epsilon=|f|/\gamma$. The value of the external force $f$ should be taken as minus the capillary force $\tilde{f}$ (Eq.\ (\ref{tildef})) acting at the particle. Indeed, due to $\tilde{\mathcal{F}}=\mathcal{F}+fh$, the equilibrium condition $\partial \mathcal{F}/\partial h|_{eq}=0$ is equivalent to $f=\partial\tilde{\mathcal{F}}/\partial h|_{eq}=\partial\tilde{F}_{A0}/\partial h=-\tilde{f}$. The capillary force $\tilde{f}(h)$ is calculated by numerical differentiation of the free energy $\tilde{F}_{A0}(h)$. Particularly, by using the relation $h(\beta)$ (Eq.\ (\ref{displacement})), one obtains the values $f/(\gamma a)=-3.56$ and $f/(\gamma a)=2.77$, for $\beta=\pi/4$ and $\beta=3\pi/4$, respectively, for which the curves $z(r)$, given implicitely by $z(\theta)=[R_0+u(\theta)]\cos\theta-z_{substr}$ and $r(\theta)=[R_0+u(\theta)]\sin\theta$ with $u(\theta)=R_0 \epsilon v_0(\theta)$ and $z_{substr}=R_0\cos\theta_0$, have been plotted in Fig.\ \ref{symmetric_analytic}.

%\bibliographystyle{plainnat}
%\bibliography{citation8}{}

\begin{thebibliography}{51}
\bibitem[{cor()}]{correction}
\bibinfo{note}{In calculating $\delta F$ we have ignored the correction $\delta
  x$, which also depends on $\alpha$ (see Eq.\ (\ref{xcmu})), but gives a
  contribution of the order $(f^2/\gamma)\times O((a/R_0)^3)$.}
\end{thebibliography}

\end{document}